\documentclass[fleqn,usenatbib]{mnras}

\usepackage{newtxtext,newtxmath}

\usepackage[T1]{fontenc}

\DeclareRobustCommand{\VAN}[3]{#2}
\let\VANthebibliography\thebibliography
\def\thebibliography{\DeclareRobustCommand{\VAN}[3]{##3}\VANthebibliography}

\usepackage{mdframed}
\usepackage{xcolor}
\usepackage{graphicx}	
\usepackage{amsmath}	
\usepackage{url}       

\title[SNID-SAGE]{SNID-SAGE: A Modern Framework for Interactive Supernova Classification and Spectral Analysis}

\author[F. Stoppa et al.]{
Fiorenzo Stoppa,$^{1}$%
\thanks{E-mail: fiorenzo.stoppa@physics.ox.ac.uk}
Stephen J. Smartt$^{1}$
\\
$^{1}$Astrophysics sub-Department, Department of Physics, University of Oxford, Denys Wilkinson Building, Keble Road, Oxford, OX1 3RH, UK
}

\date{\today}

\pubyear{\the\year{}}

\begin{document}
\label{firstpage}
\pagerange{\pageref{firstpage}--\pageref{lastpage}}
\maketitle


\begin{abstract}

We present \textsc{SNID--SAGE} (SuperNova IDentification---Spectral Analysis and Guided Exploration), a framework for supernova spectral classification with both a fully interactive graphical interface and a scriptable command-line pipeline for large-scale processing. The pipeline combines deterministic spectral preprocessing, FFT-based cross-correlation against a curated template library, ranking of candidate matches using a composite quality metric, and consolidation of redshift and classification solutions into a single result with associated quality and confidence estimates. \textsc{SNID--SAGE} includes an upgradeable template library (about 6000 spectra), interactive line identification with velocity measurements, and optional natural-language summaries of classification results.

We evaluate \textsc{SNID--SAGE} using two complementary tests: (i) leave-one-out cross-validation, in which each template spectrum is matched against the remainder of the library; and (ii) large-scale application to WISeREP spectra with valid coverage across the 4000--7000\,\AA\ interval, irrespective of spectral type, comprising approximately 46\,000 spectra, with redshift validation against known host-galaxy measurements where available. The full validation results and the \textsc{SNID--SAGE} framework are publicly available, supporting integration into spectroscopic survey workflows.

\end{abstract}


\section{Introduction}

The number and diversity of astronomical transients discovered each year are growing rapidly, driven by wide-field time-domain surveys such as Pan-STARRS \citep{flewelling2020ps1}, ATLAS \citep{tonry2018atlas}, BlackGEM \citep{groot2024blackgem}, ZTF \citep{bellm2019ztf}, and GOTO \citep{steeghs2022goto}, with further increases expected from the Rubin Observatory's Legacy Survey of Space and Time (LSST; \citealt{ivezic2019lsst}).

As discovery rates increase, the demand for spectroscopic follow-up to enable transient classification and redshift estimation grows accordingly. Classification is currently supported by a wide range of spectroscopic facilities spanning apertures from roughly 1.5 to 10\,m, with corresponding variation in spectral resolution, sensitivity, and available follow-up time. Major contributors include Keck/LRIS \citep{okeLRIS1995}, VLT/X-shooter \citep{Vernet2011}, VLT/FORS \citep{appenzellerFORS1998}, Gemini/GMOS \citep{Hook2004}, GTC/OSIRIS \citep{cepaOSIRIS1998}, NTT/EFOSC2 through PESSTO \citep{smartt2015pessto}, UH88/SNIFS \citep{tuckerSCAT2022}, Palomar/SEDM \citep{Blagorodnova2018}, LT/SPRAT \citep{Piascik2014}, NOT/ALFOSC \citep{Djupvik2010}, and LCO/FLOYDS \citep{Brown2013}. New multiplexed facilities such as WEAVE \citep{Shoko2024}, 4MOST/TiDES \citep{Dejong2019, Frohmaier2025}, and Subaru/PFS \citep{Tamura2016} will further increase the volume of available spectra, shifting the bottleneck from discovery toward scalable and reproducible classification.


Over the past two decades, a small number of community-standard tools have underpinned supernova spectral classification and redshift inference. Widely used approaches include template-based matching methods, such as those implemented in \textsc{SNID} \citep{blondin2007snid} and in $\chi^2$ fitting frameworks with host-galaxy mixing (e.g. \textsc{Superfit} and its Python re-implementation NGSF; \citealt{howell2005superfit,goldwasser2022ngsf,yaron2012wiserep}), as well as web-based spectral matching services like \textsc{GELATO} \citep{harutyunyan2008gelato} and machine-learning classifiers such as \textsc{DASH} \citep{muthukrishna2019dash}. These tools have proven highly effective in traditional follow-up settings; however, the scale, cadence, and heterogeneity of current and forthcoming spectroscopic programmes place new demands on classification pipelines, requiring methods that remain robust across diverse transient classes and readily extensible to incorporate new or rare types without retraining or reconfiguration.

We introduce \textsc{SNID--SAGE}, a spectroscopic classification framework that retains the template-based cross-correlation pioneered by \textsc{SNID} \citep{tonry1979, blondin2007snid} and augments it with a new core metric and a redshift-space Gaussian mixture model (GMM) clustering to combine information from multiple template matches rather than relying on a single best-matching template.
\textsc{SNID--SAGE} is provided as both a fully interactive graphical interface and a scriptable command-line pipeline. In addition to classification, it integrates tools for template curation, emission- and absorption-line analysis, and natural-language summaries, the latter remaining strictly descriptive and independent of the numerical inference.

This paper is organised as follows. Section~\ref{sec:methods} describes the \textsc{SNID--SAGE} classification pipeline, including preprocessing, cross-correlation, similarity metrics, and clustering-based inference. Section~\ref{sec:templates} outlines the construction and composition of the template library. Section~\ref{sec:validation} presents validation and performance analyses based on leave-one-out testing and large-scale application to WISeREP spectra. Finally, Section~\ref{sec:conclusions} summarises the main results and discusses limitations and future directions.

\section{Methods}
\label{sec:methods}

\textsc{SNID--SAGE} implements a deterministic spectroscopic classification pipeline with two interfaces, an interactive graphical client and a batch-oriented command-line tool. At a high level, the pipeline ingests a wavelength-calibrated one-dimensional spectrum, applies a fixed preprocessing sequence, compares the conditioned spectrum against a library of templates, and consolidates the resulting ensemble of matches into a type, phase, and redshift inference. Alongside the primary classification, the pipeline produces diagnostic outputs and alternative solutions to support interpretation and quality control.

In this section, we describe the preprocessing, cross-correlation, similarity metrics, and clustering-based consolidation steps used to derive type, redshift, and phase (days relative to maximum light) estimates from an observed spectrum.

\subsection{Step 1: Preprocessing}
\label{sec:methods:preproc}

Preprocessing follows the principles of the classical \textsc{SNID} approach \citep{blondin2007snid}, transforming an observed spectrum into a preprocessed representation suitable for FFT-based cross-correlation. 

\begin{figure}
\centering 
\includegraphics[width=\linewidth]{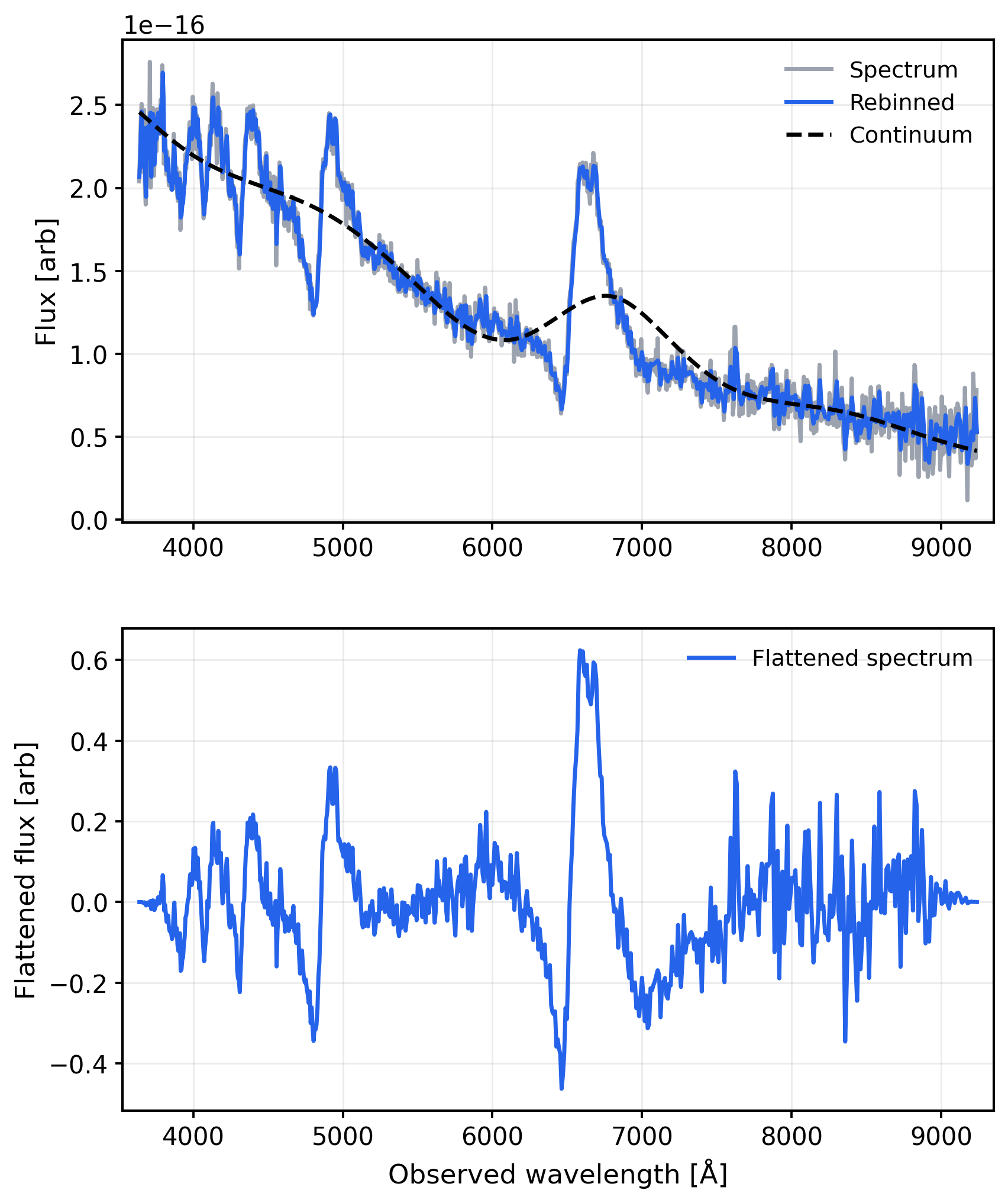} 
\caption{\textbf{Preprocessing pipeline} illustrated for SN\,2018bif. 
\textit{Top panel:} Raw input spectrum (grey) and spectrum after logarithmic-wavelength rebinning (blue), with the fitted broadband continuum overlaid (dashed black). 
\textit{Bottom panel:} Continuum-divided flattened spectrum ready for the cross-correlation stage.}
\label{fig:preproc_pipeline} 
\end{figure}

The first stage addresses data quality. Obvious artefacts such as cosmic-ray spikes and isolated outliers are identified automatically and masked prior to further processing. The user may additionally define wavelength regions to exclude, such as telluric absorption bands (e.g. the A-band at 7550--7700\,\AA) or known instrumental artefacts. Optional Savitzky--Golay smoothing \citep{savitzky1964} can be applied to reduce high-frequency noise while preserving spectral features.
In the second stage, the observed spectrum is rebinned onto the default logarithmic wavelength grid used by the template library. The templates are pre-defined on a common $\ln\lambda$ grid spanning 2500--10000\,\AA\ with 1024 uniformly spaced bins, and the input spectrum is interpolated onto this same grid prior to cross-correlation. Working in $\ln\lambda$ ensures that a cosmological redshift $z$ corresponds to an additive shift $\Delta\ln\lambda = \ln(1+z)$; on a grid with spacing $\Delta\ln\lambda$, this is a shift of $\Delta k = \ln(1+z)/\Delta\ln\lambda$ bins in correlation space, enabling efficient FFT-based matching \citep{tonry1979}.
The third stage removes the broadband continuum. A low-order spline model is fitted to the log-rebinned spectrum and divided out, isolating the line structure that drives the correlation and reducing sensitivity to broadband differences between the target and templates, consistent with the established \textsc{SNID} continuum treatment \citep{blondin2007snid}. After continuum division, unity is subtracted to centre the normalised continuum at zero, and the spectrum is then tapered at the edges with a cosine window to suppress discontinuities that would otherwise introduce spurious high-frequency power in the Fourier domain \citep{harris1978}. The result is a zero-mean, flattened spectrum on a regular $\ln\lambda$ grid, ready for the cross-correlation stage. Figure~\ref{fig:preproc_pipeline} illustrates the input spectrum and its final preprocessed form. Although the preprocessing is generally well behaved, continuum fitting can, in principle, modify very broad spectral features; however, the use of a low-order spline and its consistent application to both target and template spectra helps ensure that residual distortions are treated consistently between target and template spectra, reducing their impact on the cross-correlation.

\subsection{Step 2: Cross-correlation and similarity metrics}
\label{sec:methods:fft}

After preprocessing, the flattened target spectrum is compared with each rest-frame template using the Fourier cross-correlation technique introduced by \citet{tonry1979} and adopted by \textsc{SNID} \citep{blondin2007snid}. Working on the common $\ln\lambda$ grid described in the previous section, redshift determination reduces to identifying the shift that maximises the alignment of spectral features between the target and each template.

\begin{figure}
    \centering
    \includegraphics[width=\linewidth]{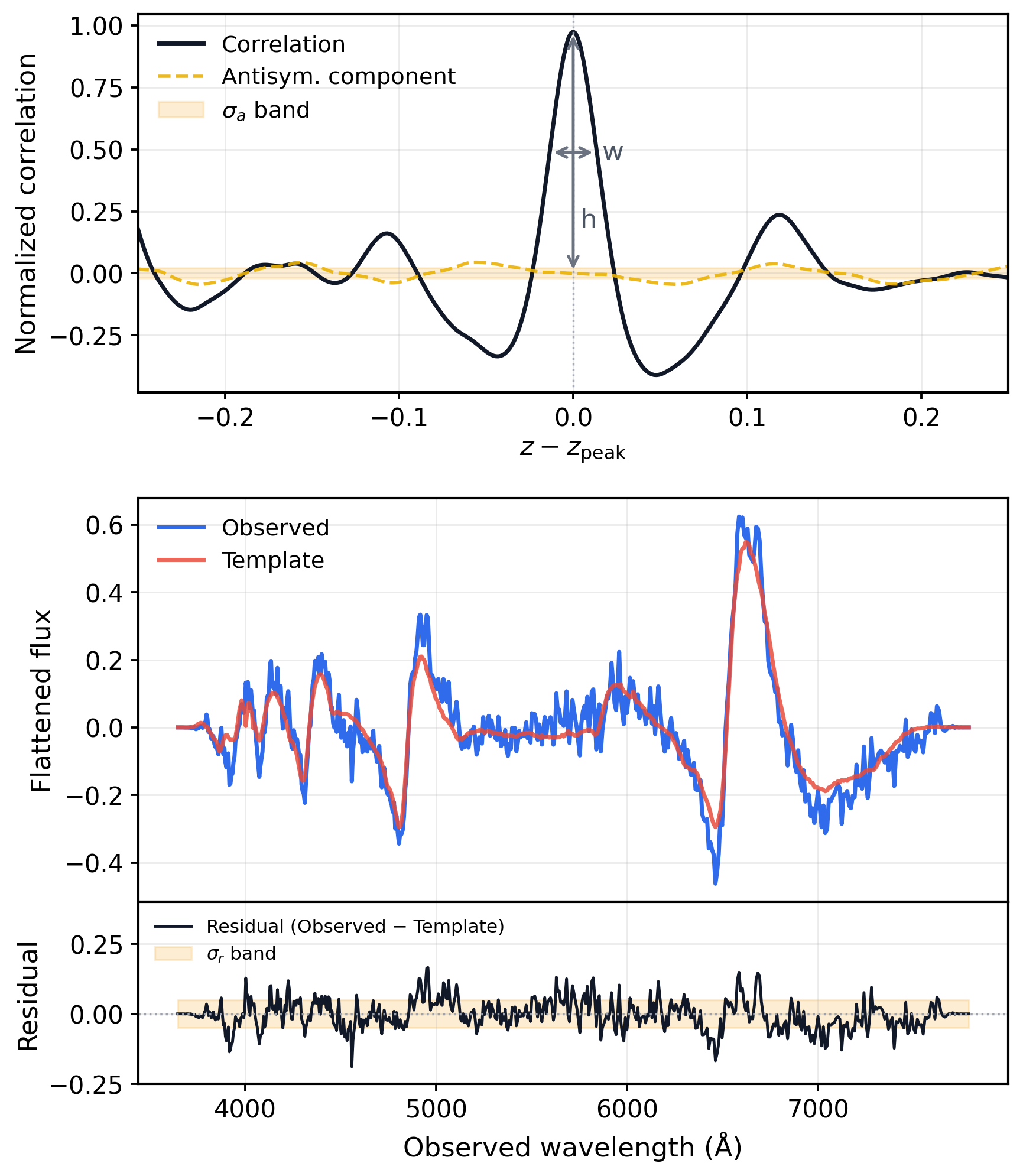}
    \caption{\textbf{Illustration of cross-correlation components and match-quality diagnostics for SN\,2018bif.}
    \textit{Top panel:} Normalised cross-correlation function between the observed spectrum and the best-matching template, shown as a function of redshift offset relative to the peak ($z - z_{\rm peak}$). The solid black curve shows the filtered correlation, while the dashed orange curve shows the antisymmetric component used in the classical Tonry--Davis $R$ statistic. The shaded band indicates the rms of the antisymmetric component. The peak height h and width w characterise the strength of the alignment and the sharpness of the redshift constraint.
    \textit{Middle panel:} Continuum-flattened, apodised observed spectrum (blue) and the best-matching template (red) at the peak redshift, illustrating the degree of spectral agreement that drives the correlation.
    \textit{Bottom panel:} Residual spectrum (observed minus template) at the peak redshift, with the shaded band indicating the standard deviation of the residuals. In \textsc{SNID--SAGE}, this residual scatter provides an explicit estimate of mismatch, complementing the peak width in the redshift uncertainty estimate and replacing reliance on the antisymmetric correlation component alone.}
    \label{fig:crosscorr_diagnostics}
\end{figure}

To suppress spurious correlations driven by residual continuum structure or high-frequency noise, the cross-correlation is computed with a smooth raised-cosine band-pass filter in Fourier space, following standard practice \citep{harris1978,blondin2007snid}. The resulting cross-correlation function typically exhibits one or more peaks, each corresponding to a candidate redshift (Figure~\ref{fig:crosscorr_diagnostics}, top panel). For a given template, we only take the highest peak as the preferred redshift solution, and its centre position is further refined via local quadratic interpolation to achieve sub-grid accuracy.

In the original \textsc{SNID} formulation, the significance of a cross-correlation peak is quantified using the Tonry--Davis $R$ statistic, defined as the peak height divided by the rms of the antisymmetric component of the correlation function ( $R = h \; / (\sqrt{2} \sigma_a)$, components shown in Figure~\ref{fig:crosscorr_diagnostics}). This is combined with the fractional wavelength overlap between the target and the redshifted template to form the standard quality metric $\mathrm{RLAP}=R\times\mathrm{lap}$ \citep{blondin2007snid}. As illustrated in the top panel of Figure~\ref{fig:crosscorr_diagnostics}, the antisymmetric component is intended to provide a noise estimate against which the peak height is normalised. However, the antisymmetric component of the cross-correlation is not, in general, a reliable estimator of the effective noise in the match. As noted in several studies, it can be influenced by spectral structure, windowing effects, and filtering choices rather than noise alone \citep{Heavens1993,Bouchy2001,Zucker2003,Zamora2023}. As a result, $\mathrm{RLAP}$ can overestimate the quality of matches in cases where the correlation peak is narrow but driven by a limited subset of features.
Furthermore, cross-correlation is intrinsically insensitive to overall amplitude and offset mismatches, as it quantifies relative alignment but not agreement in flux scale.

In \textsc{SNID--SAGE}, we aim to address these two limitations directly. 
First, rather than using the antisymmetric rms as a proxy for noise, we characterise peak sharpness and spectral mismatch explicitly. For each template, we define a local scale parameter
\[
\sigma = w\,\sigma_r,
\]
where $w$ is the full width at half maximum (FWHM) of the cross-correlation peak in redshift space and $\sigma_r$ is the standard deviation of the residual spectrum (observed minus template) evaluated at the peak redshift (Figure~\ref{fig:crosscorr_diagnostics}, bottom panel). This scale increases for broad or poorly matched solutions and moderates the ranking accordingly. Second, to account for amplitude and offset mismatches that cross-correlation alone cannot capture, we incorporate the Lin concordance correlation coefficient \citep{lin1989}, $\mathrm{CCC}$, which measures agreement in both correlation and flux scaling between the target and the redshifted template.

Template matches in \textsc{SNID--SAGE} are therefore ranked using a composite statistic
\[
\mathrm{H\sigma LAP\text{-}CCC}
=
\frac{h \; \mathrm{LAP} \; \mathrm{CCC}}{\sqrt{\sigma}},
\]
where $h$ is the cross-correlation peak height, $\mathrm{LAP}$ the fractional wavelength overlap, $\mathrm{CCC}$ the concordance correlation coefficient, and $\sigma$ the local sharpness–mismatch scale defined above. The square-root dependence on $\sigma$ introduces a moderate penalty for broad or spectrally inconsistent solutions while preserving the primary influence of the alignment-quality terms in the numerator. As with $\mathrm{RLAP}$, this statistic is heuristic and intended for stable ranking rather than direct probabilistic interpretation.

For each template, the procedure therefore yields a best-fitting redshift, the corresponding match score $\mathrm{H\sigma LAP\text{-}CCC}$, and associated metadata (type, subtype, and phase). Templates with $\mathrm{H\sigma LAP\text{-}CCC} \le 1.5$ are discarded at this stage, so that only meaningful matches are propagated to the consolidation step.


\subsection{Step 3: Clustering and inference}
\label{sec:methods:gmm}

In the original SNID framework, classification is based on the ranking of individual template matches, typically using the $RLAP$ statistic, and the final type and redshift are inferred from the highest-scoring templates or from simple aggregation over the top-ranked subset \citep{blondin2007snid}. While effective in many cases, this approach treats template matches largely independently and does not explicitly account for the presence of multiple, potentially distinct solutions in redshift space.
\textsc{SNID--SAGE} instead combines information from multiple template matches in a structured way to derive a more stable and interpretable classification.

\begin{figure}
    \centering
    \includegraphics[width=1\linewidth]{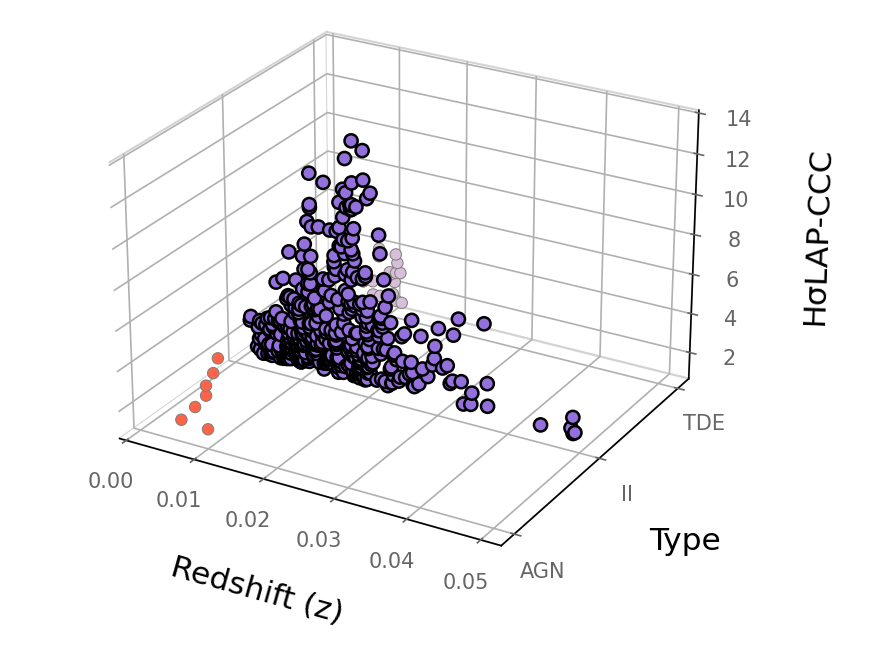}
    \caption{\textbf{Example of redshift-space clustering and cluster selection in \textsc{SNID--SAGE}.}
    Visualisation of template matches for 2018bif shown as a function of redshift, template type, and $H\sigma\mathrm{LAP\text{-}CCC}$. Each point corresponds to a single template match; clusters form naturally as concentrations of matches at similar redshift within a given type. The highlighted cluster corresponds to the highest-ranked solution according to the cluster quality score $Q$.
    }
    \label{fig:clustering_example}
\end{figure}

For each template type (currently 15 categories: AGN, CV, GAP, Galaxy, II, Ia, Ib, Ibn, Ic, Icn, KN, LFBOT, SLSN, Star, and TDE), we collect the redshift estimates of the retained template matches and cluster them using a one-dimensional Gaussian mixture model (GMM; \citealt{McLachlan2000}). This procedure yields distinct redshift clusters that represent alternative candidate solutions within a given type. The number of Gaussian components is selected using the Bayesian Information Criterion (BIC; \citealt{schwarz1978bic}), supplemented by an elbow criterion to avoid over-fragmentation in sparsely populated regions of redshift space. To ensure physically meaningful solutions, clusters are required to correspond to contiguous structures in redshift space: widely separated groups of matches are split into distinct clusters rather than merged into a single solution. An example of this clustering and subsequent cluster selection is shown in Figure~\ref{fig:clustering_example}.

For each cluster, we define a quality score $Q$ based on its highest-ranked member templates. Specifically, we compute the mean of the top five $H\sigma\mathrm{LAP\text{-}CCC}$ scores within the cluster (or all members if fewer than five are present) and apply a linear penalty when fewer than five templates support the solution:
\[
Q = \left(\frac{n}{5}\right)\,\overline{q}_{\rm top},
\qquad n = \min(5, N_{\rm cl}),
\]
where $\overline{q}_{\rm top}$ is the mean of the top $n$ match $H\sigma\mathrm{LAP\text{-}CCC}$ scores and $N_{\rm cl}$ is the cluster size. 
This definition favours solutions that are both strong and supported by multiple independent templates.

Across all clusters and types, the cluster with the highest $Q$ is adopted as the primary solution, while lower-ranked clusters are retained as explicit alternatives. For reporting, $Q$ is mapped to qualitative classes using fixed thresholds: \emph{Very Low} for $Q<2.5$, \emph{Low} for $2.5\le Q<5$, \emph{Medium} for $5\le Q<8$, and \emph{High} for $Q\ge 8$. 

We additionally compute a relative type confidence based on the fractional improvement in $Q$ over the next-best competing type,
\[
C_{\rm type} = \frac{Q_1 - Q_2}{Q_2}\times 100\%,
\]
where $Q_1$ is the winning type’s best-cluster score and $Q_2$ is the best-cluster score of the runner-up type. This is mapped to discrete confidence classes using fixed thresholds: \emph{High} for $C_{\rm type}\ge 75\%$, \emph{Medium} for $C_{\rm type}\ge 25\%$, \emph{Low} for $C_{\rm type}\ge 5\%$, and \emph{Very Low} otherwise. When a runner-up type is absent after the match-retention and clustering steps, no finite value of $C_{\rm type}$ is assigned. These cases are flagged separately as \emph{No Comp}, indicating that no competing alternative survived the quality criteria.

Each redshift cluster typically contains template matches belonging to multiple subtypes of the same type (see Table\,\ref{tab:template_breakdown} for the lists of subtypes). 
Within the highest-ranked cluster, subtype inference is performed by grouping member templates by subtype and applying the same top-$n$ scoring procedure to each subgroup. The subtype with the highest resulting $Q$ is selected as the preferred subtype, and a relative subtype confidence is computed from the fractional improvement over the next-best subtype within the same cluster.

With the preferred subtype established, we estimate its redshift and phase by combining information from all templates of that subtype within the winning cluster. For each contributing template $i$, we retain the redshift $z_i$, phase $t_i$, and match score $q_i$, where $q_i$ is the template’s $H\sigma\mathrm{LAP\text{-}CCC}$ value (Section~\ref{sec:methods:fft}).
Templates are weighted according to their relative match quality using exponential weights,
\[
w_i = \exp\!\left(\frac{q_i}{\tau}\right),
\]
which favours higher-quality matches while retaining contributions from the full subtype sample. In practice, we adopt $\tau=1$, which provides stable behaviour across the template library. The parameter $\tau$ controls the concentration of the weights: when one match is clearly superior, it dominates the estimate, whereas comparable matches contribute collectively when scores are similar. This weighting is heuristic and designed for robust aggregation rather than formal probabilistic inference.

We then estimate redshift and phase as weighted means,
\[
\hat{z} = \frac{\sum_i w_i z_i}{\sum_i w_i} \, , \qquad
\hat{t} = \frac{\sum_i w_i t_i}{\sum_i w_i} \, ,
\]
and report internal-scatter measures defined as the unbiased weighted dispersion,
\[
\sigma_{\hat{x}} = \left[
  \frac{\sum_i w_i (x_i - \hat{x})^2}{\sum_i w_i}
  \,\times\,
  \frac{N_\mathrm{eff}}{N_\mathrm{eff}-1}
\right]^{1/2},
\]
where $x=z$ (with $x_i=z_i$) gives $\sigma_{\hat{z}}$ and $x=t$ (with $x_i=t_i$) gives $\sigma_{\hat{t}}$, and where
\[
N_\mathrm{eff} = \frac{\left(\sum_i w_i\right)^2}{\sum_i w_i^2} \, .
\]

\noindent
This correction intentionally penalises solutions that are dominated by only one or a few templates: when a single match carries most of the weight, $N_\mathrm{eff}\rightarrow 1$ and the factor $N_\mathrm{eff}/(N_\mathrm{eff}-1)$ increases the reported scatter, reflecting the fact that a visually strong best match is less reliable when it is not supported by a broader set of comparably good templates. In Section~\ref{sec:validation:loo}, we validate these redshift and phase estimators using leave-one-out cross-validation, including residual and uncertainty calibration analyses.

At the end of this stage, the core \textsc{SNID--SAGE} analysis is complete. The pipeline returns a primary classification consisting of a best-fitting type and subtype, together with redshift and phase estimates and their associated internal-scatter measures. These results are accompanied by an absolute quality assessment for the winning solution and relative confidence measures that quantify how decisively the preferred type and subtype are favoured over competing alternatives. Lower-ranked clusters and subtype solutions are retained as explicit alternatives, enabling easy inspection and downstream decision-making by the user.

\subsection{Additional GUI Features}
\label{sec:methods:gui-features}

In addition to the core classification pipeline described above, the \textsc{SNID--SAGE} graphical interface provides a set of auxiliary analysis tools designed to support inspection, refinement, and interpretation of spectroscopic classifications. These tools operate on the results of the deterministic pipeline and do not modify the underlying classification logic.

\subsubsection{Host redshift constraints}
\label{sec:methods:host-redshift}

When a host-galaxy redshift is available from independent observations (e.g.\ narrow emission lines), it can be used to constrain the analysis. In the graphical interface, the host redshift may be either fixed to a single value, in which case the cross-correlation is evaluated only at that redshift, or specified as an interval, in which case the search is restricted to that redshift range (default $\pm 0.0005$ about the chosen value). The interface supports both manual identification of common spectral features (e.g.\ H$\alpha$, H$\beta$, [O~III]) and automatic estimation via galaxy template matching.

The scripted workflow provides the same functionality via the \texttt{--forced-redshift} option: if a fixed value is supplied, the templates are matched only at that redshift; if a bounded interval is supplied, the cross-correlation search is limited to that interval, with the same default width of $\pm 0.0005$ when no custom range is specified. In both cases, constraining the redshift in this way reduces the search space and can improve classification robustness and phase estimation when the host redshift is well determined.

\subsubsection{Line identification and velocity measurements}
\label{sec:methods:gui-lines}

Given a consolidated redshift estimate, the \textsc{SNID--SAGE} GUI provides an interactive line-identification tool to aid inspection of both supernova and host-galaxy features. Users can overlay curated rest-frame line lists (selectable by transient type, approximate phase, and element group) and interactively toggle individual lines to visualise their expected observed wavelengths.

For any selected line, the GUI supports a lightweight measurement workflow: the user marks points around the feature and the tool returns basic line properties such as the peak wavelength and a FWHM estimate. When a rest wavelength is available, the measured width is converted into a characteristic velocity scale. These measurements are intended to support rapid, transparent interpretation of spectra (e.g.\ checking consistency with a proposed type/phase or identifying host emission lines), and do not feed back into the automated classification.

\subsubsection{AI-powered analysis summaries}
\label{sec:methods:gui-ai}

The GUI includes an optional \emph{AI Assistant} button that generates a short, structured natural-language summary of the current \textsc{SNID--SAGE} results. When enabled, the tool formats the classification output (best type/subtype, redshift and phase estimates, match quality/confidence, and any user-selected line markers) together with optional observation metadata entered by the user, and submits this context to an externally configured LLM service (via OpenRouter). The returned text is intended for rapid reporting and human-readable logging.

The AI Assistant is strictly descriptive: it does not modify preprocessing, matching, clustering, or any numerical outputs, and its summary should not be interpreted as an independent classifier. Because this feature relies on an externally configured third-party service, it is disabled unless explicitly set up by the user, and it should be used with appropriate caution for proprietary or unpublished data.

\section{Template Library}
\label{sec:templates}

The performance of \textsc{SNID--SAGE} is fundamentally determined by the quality, diversity, and accuracy of its spectral template library. The current default optical library comprises 698 rest-frame templates (5999 individual spectra) rebinned onto a common logarithmic wavelength grid (2500--10000\,\AA; 1024 bins in $\ln\lambda$). 

The majority of templates originate from the public SuperSNID library \citet{Magill2025}, supplemented by stripped-envelope templates (Ib, Ic, Ibn, Icn) created by the Modjaz METAL group \citet{Yesmin2024, Williamson2023, Williamson2019, Liu2017, Modjaz2016, Liu2016, Liu2014}, together with additional templates curated and validated as part of this work. Table~\ref{tab:template_breakdown} summarises the template types and subtypes represented in the default optical library.

\begin{table*}
\centering
\caption{Composition of the 698-template default optical library, grouped by transient type. Subtype counts are shown in parentheses. The total number of individual spectra (epochs) per type is given in the final column.}
\label{tab:template_breakdown}
\begin{tabular}{lccc}
\hline
Type & Subtypes (N) & Templates & Spectra \\
\hline
Ia &
\begin{tabular}[t]{@{}l@{}}
Ia-norm (160), Ia-91T (36), Ia-91bg (33), Ia-02cx (12), Ia-csm (10) \\
Ia-02es (7), Ia-03fg (7), Ia-99aa (2), Ia-Ca-rich (2), Ia-pec (2)
\end{tabular}
& 271 & 2864 \\

II &
\begin{tabular}[t]{@{}l@{}}
IIP (49), IIb (31), IIn (16), II-flash (5), IIL (4), \\
II-87A (1), II-pec (1), IIn-pec (1)
\end{tabular}
& 108 & 1216 \\

Ib &
Ib-norm (44), Ib-pec (6), Ib-Ca-rich (4)
& 54 & 509 \\

Ic &
Ic-norm (36), Ic-broad (30), Ic-05ek (2), Ic-Ca-rich (1)
& 69 & 632 \\

Ibn & Ibn (12) & 12 & 73 \\
Icn & Icn (5) & 5 & 29 \\
SLSN & SLSN-I (48), SLSN-II (13) & 61 & 297 \\
TDE & TDE-H-He (11), TDE-He (5), TDE-H (3), TDE-Ftless (1) & 20 & 105 \\
AGN & QSO (4), AGN-Sey2 (2), QSO2 (2), AGN-type1 (1), QSO-BAL (1), Seyfert 1.5 (1) & 11 & 15 \\
Galaxy & Gal-SB (12), Gal-Sc (5), Gal-E (4), Gal-S0 (2), Gal-Sa (2), Gal-Sb (2) & 27 & 27 \\
CV & DN (4), Nova (4), NL (2), AM\_CVn (1), IP (1), Polar (1) & 13 & 68 \\
GAP & LBV (5), ILRT (5), LRN (4) & 14 & 72 \\
LFBOT & 20xnd (2), 18cow (1) & 3 & 21 \\
KN & 17gfo (1) & 1 & 10 \\
Star & Varstar (11), Mdwarf (9), Symbiotic (3), WR-WC (3), WR-WN (3) & 29 & 61 \\
\hline
Total & & 698 & 5999 \\
\hline
\end{tabular}
\end{table*}

Because template-based classification inherits any biases or inaccuracies present in the library, template quality control is essential to the reliability of the pipeline. We actively maintain and update the library and encourage community feedback. Researchers who identify errors in template metadata (e.g.\ redshift, type/subtype, or phase) or who wish to contribute additional well-characterised spectra are invited to contact the authors; proposed additions are reviewed and, when validated, incorporated into future releases of the default library.

In addition to the default optical library, an extended optical+near-infrared (ONIR) template library is currently under development and is already available in a preview form for testing. This extended library incorporates near-infrared spectral coverage, enabling cross-correlation over a broader wavelength baseline and therefore extending the effective redshift range over which reliable matches can be obtained.

\section{Validation and Performance Analysis}
\label{sec:validation}

We evaluate \textsc{SNID--SAGE} through two complementary validation approaches: (i) leave-one-out cross-validation on the template library to quantify type-classification accuracy and redshift and phase recovery; and (ii) large-scale application to the WISeREP database \citep{yaron2012wiserep}, with comparison to independently measured host-galaxy redshifts, to assess redshift accuracy, precision, and robustness under realistic survey conditions.

\subsection{Leave-One-Out Cross-Validation}
\label{sec:validation:loo}

To quantify type-classification accuracy and the reliability of redshift and phase estimation within the template library, we perform leave-one-out cross-validation on the full set of templates. In this experiment, all spectra belonging to a template are removed in turn and analysed against the remaining library using the complete \textsc{SNID--SAGE} pipeline. This procedure tests whether a spectrum can be correctly classified and assigned consistent redshift and phase estimates when its exact template is excluded, providing a good assessment of template coverage and the robustness of the clustering-based consolidation. Table~\ref{tab:loo_results_by_type} summarises per-type recovery for top-1, top-2 and top-3 predicted types together with the median absolute redshift error for correctly recovered spectra. Phase errors are reported only for transient classes with a well-defined maximum-light reference; for non-transient classes (e.g.\ galaxies, AGN, stars, CVs), phase is not defined and is therefore omitted. For example, the ``top-3'' recovery column gives the fraction of spectra for which the correct type appears among the three highest-ranked final solutions returned by \textsc{SNID--SAGE}, with ranking determined by the cluster-level quality score $Q$. Here, we consider type recovery only; redshift and phase consistency are assessed separately in the analyses that follow.

\begin{table}
\centering
\caption{Leave-one-out validation results by transient type. Reported quantities are top-$k$ type recovery (\%), median absolute redshift error $|\Delta z|$, and median absolute phase error $|\Delta t|$ (days). Classes with too few templates for statistically meaningful leave-one-out evaluation (e.g.\ KN and LFBOT) are omitted.}
\label{tab:loo_results_by_type}
\begin{tabular}{lrrrrr}
\hline
Type & Top-1 (\%) & Top-2 (\%) & Top-3 (\%) & $|\Delta z|$ & $|\Delta t|$ \\
\hline
Ia & 99.3 & 99.8 & 99.9 & $0.00146$ & $1.9$ \\
II & 88.6 & 97.9 & 99.3 & $0.00158$ & $10.9$ \\
Ibn & 82.2 & 87.7 & 90.4 & $0.00146$ & $8.9$ \\
Ic & 80.1 & 95.6 & 98.0 & $0.00349$ & $6.6$ \\
Ib & 78.5 & 94.1 & 97.6 & $0.00352$ & $6.2$ \\
GAP & 77.1 & 90.0 & 90.0 & $0.00044$ & $17.6$ \\
TDE & 65.3 & 95.8 & 96.8 & $0.00108$ & $41.0$ \\
SLSN & 56.1 & 74.8 & 81.7 & $0.00336$ & $11.5$ \\
Icn & 37.0 & 48.1 & 48.1 & $0.00111$ & $3.4$ \\
AGN & 100.0 & 100.0 & 100.0 & $0.00008$ & --- \\
Star & 91.7 & 95.0 & 95.0 & $0.00041$ & --- \\
Galaxy & 90.5 & 100.0 & 100.0 & $0.00007$ & --- \\
CV & 86.5 & 100.0 & 100.0 & $0.00022$ & --- \\
\hline
\end{tabular}
\end{table}

Overall, the per-type breakdown in Table~\ref{tab:loo_results_by_type} shows that \textsc{SNID--SAGE} recovers the correct type reliably across the majority of classes. While the top-1 recovery varies depending on spectral diversity and template coverage, the top-2 and top-3 recovery rates are consistently much higher, indicating that the true type is very frequently contained within the first few candidate solutions returned by the pipeline. 
The median redshift errors remain small for the major classes, while the phase errors are generally modest but more sensitive to template coverage and intrinsic spectral variability.

Type~Ia spectra are recovered with near-perfect top-1 accuracy (99.3\%), reflecting both their distinctive spectral morphology and the extensive phase coverage of the Ia template set. Type~II supernovae also exhibit strong recovery performance (top-1 $\sim$89\%), while stripped-envelope subclasses (Ib and Ic) show moderately lower top-1 rates (78--80\%) but substantially improved top-2 and top-3 recall, consistent with their greater spectral diversity and overlap between neighbouring subtypes at certain phases. Classes such as SLSNe, TDEs, gap transients, and Icn exhibit reduced top-1 recovery, which reflects both intrinsic heterogeneity and comparatively smaller template coverage; for many of these classes, the correct type nonetheless appears frequently within the top-2 or top-3 candidates. Very small classes (e.g.\ kilonova, LFBOT) are omitted from this table when leave-one-out evaluation is statistically unstable (removing a single template can eliminate the only representative of that class from the library), preventing meaningful recovery assessment.

\begin{figure}
\centering
\includegraphics[width=\linewidth]{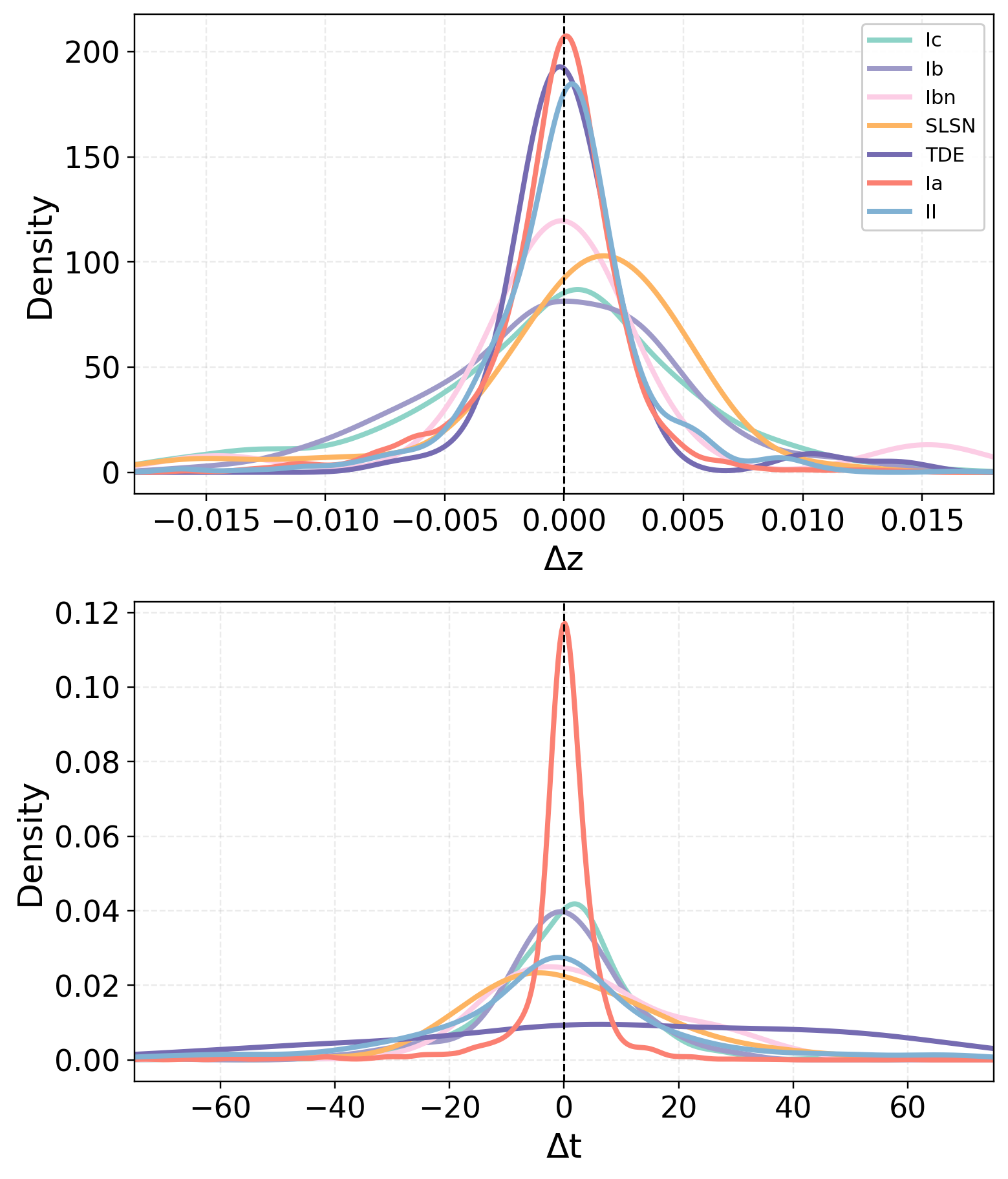}
\caption{\textbf{Leave-one-out residual distributions by transient type.}
Top: distribution of redshift residuals $\Delta z = z_{\mathrm{pred}} - z_{\mathrm{true}}$.
Bottom: distribution of phase residuals $\Delta t = t_{\mathrm{pred}} - t_{\mathrm{true}}$.
Vertical dashed lines mark zero residual. Redshift residuals are tightly centred for all major classes, while phase residuals show a stronger dependence on transient type, with broader distributions and extended tails for heterogeneous classes such as SLSNe and TDEs.}
\label{fig:loo_residuals}
\end{figure}

In addition to overall recovery rates, we examine the distribution of redshift and phase residuals obtained in the leave-one-out experiment. 
Figure \ref{fig:loo_residuals} shows kernel-density estimates of $\Delta z$ and $\Delta t$ (predicted minus true values) for a representative subset of transient classes.
The redshift residuals are tightly centred around zero for all major supernova classes, with Type Ia exhibiting the narrowest distribution. Core-collapse classes show broader wings, reflecting greater spectral diversity and increased degeneracy between neighbouring types. Phase residuals exhibit a stronger dependence on transient class, with Type Ia again showing the narrowest distribution, while classes such as SLSNe and TDEs display substantially broader and more extended residuals. This reflects both the more heterogeneous spectral evolution of these populations and the more limited template coverage. In particular, the presence of extended tails in the phase residuals indicates that phase estimates for these classes can be significantly less constrained.

\begin{figure}
\centering
\includegraphics[width=\linewidth]{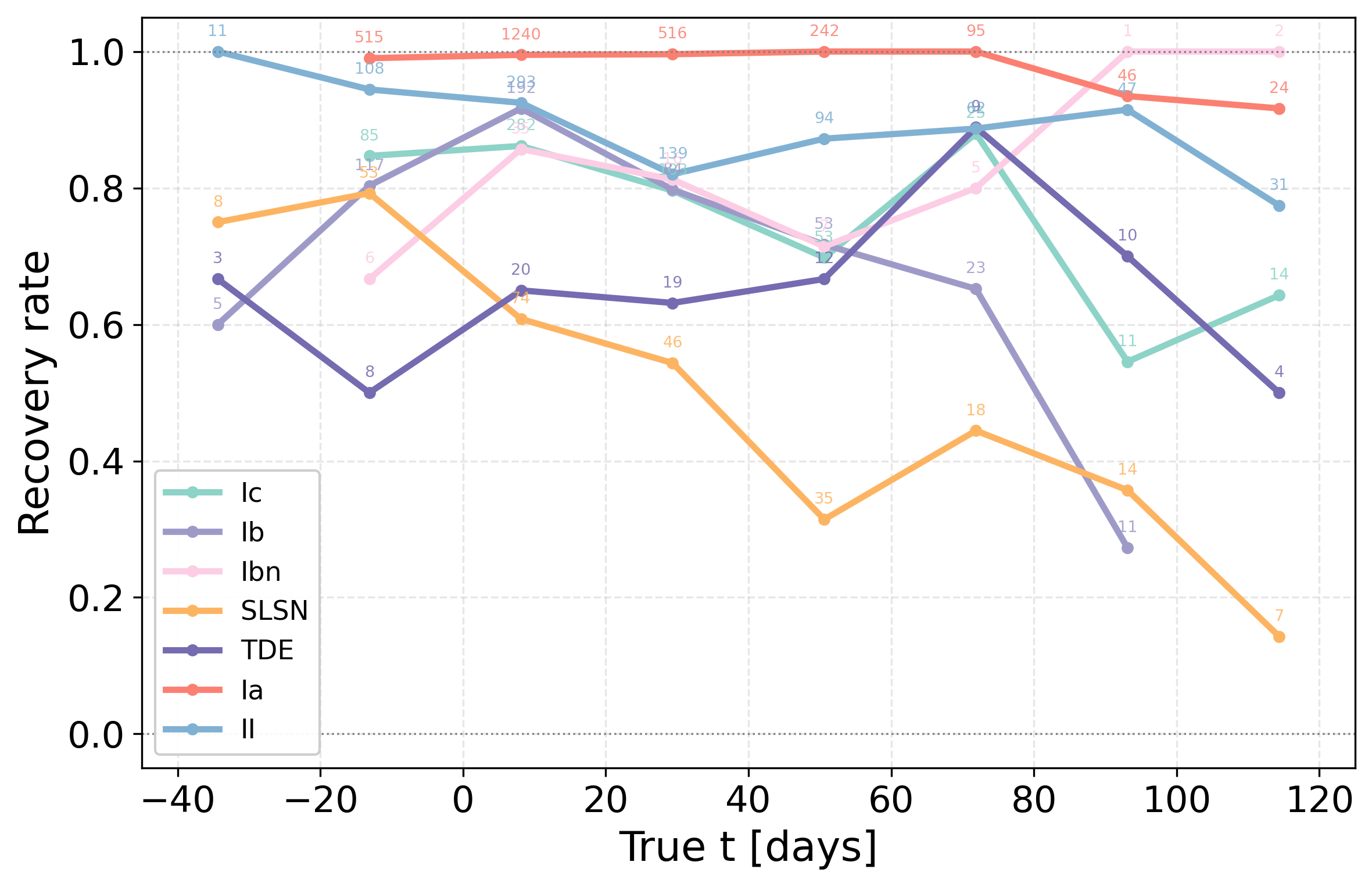}
\caption{\textbf{Leave-one-out recovery rate as a function of phase for selected transient classes.} Classification performance is highest near peak light and degrades at late phases, with stronger phase dependence and variability for heterogeneous and sparsely sampled classes such as SLSNe and TDEs.}
\label{fig:loo_epoch_recovery}
\end{figure}

To explore how classification performance depends on evolutionary phase, we compute the recovery rate as a function of phase for a representative subset of transient types. Figure~\ref{fig:loo_epoch_recovery} shows the fraction of correctly recovered spectra in phase bins.
For Type~Ia supernovae, recovery remains near unity across all sampled phases, reflecting both strong spectral homogeneity and dense template coverage. Type~II supernovae also exhibit consistently high recovery, with only modest degradation at later phases as spectral features evolve. Stripped-envelope classes (Ib and Ic) show a stronger dependence on phase, with recovery declining at late times where line blending and signal-to-noise limitations become more pronounced. SLSNe and TDEs display both lower overall recovery and significantly larger variability with phase, including pronounced degradation at intermediate and late phases.
Overall, these trends indicate that classification robustness is strongest near peak light and becomes increasingly sensitive to intrinsic diversity and template coverage at extreme early or late phases.

To assess whether the internal uncertainty estimates produced by \textsc{SNID--SAGE} are well calibrated, we examine the distribution of normalised residuals obtained in the leave-one-out experiment. For each correctly classified spectrum we compute $(z_{\mathrm{pred}}-z_{\mathrm{true}})/\sigma_z$ and $(t_{\mathrm{pred}}-t_{\mathrm{true}})/\sigma_t$, where $\sigma_z$ and $\sigma_t$ are the internal uncertainty estimates reported by the pipeline. If the uncertainties are well calibrated and the residuals are approximately Gaussian, these normalised residuals are expected to follow a standard normal distribution $\mathcal{N}(0,1)$.

\begin{figure}
\centering
\includegraphics[width=\linewidth]{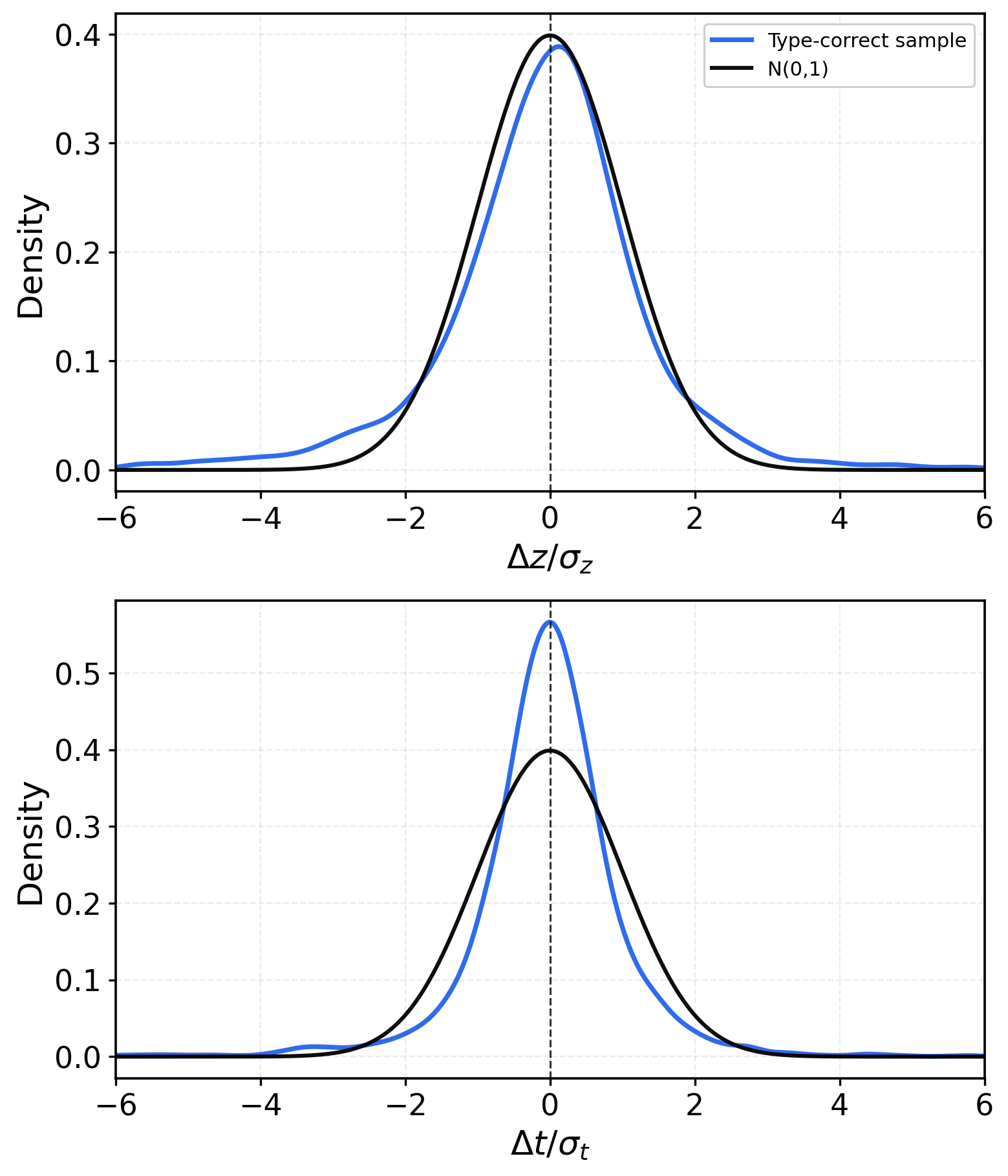}
\caption{\textbf{Calibration of internal uncertainty estimates from the leave-one-out experiment.}
Top: distribution of normalised redshift residuals $(z_{\mathrm{pred}}-z_{\mathrm{true}})/\sigma_z$ for spectra with correctly recovered type. 
Bottom: distribution of normalised phase residuals $(t_{\mathrm{pred}}-t_{\mathrm{true}})/\sigma_t$. 
In both panels, the blue curve shows the kernel-density estimate of the empirical distribution, while the black curve shows the unit normal distribution $\mathcal{N}(0,1)$ expected for well-calibrated uncertainties. The vertical dashed line marks zero residual.}
\label{fig:loo_pull_calibration}
\end{figure}

Figure~\ref{fig:loo_pull_calibration} compares the empirical distributions of the normalised residuals with the unit normal expectation. The redshift residuals closely follow the $\mathcal{N}(0,1)$ distribution, indicating that the redshift uncertainties derived from the cluster-weighted solution provide a realistic estimate of the true scatter. In contrast, the phase residuals are more narrowly distributed than a unit normal, implying that the reported phase uncertainties are somewhat conservative. In other words, the pipeline tends to slightly overestimate the uncertainty on phase estimates, such that the true phase errors are typically smaller than the nominal uncertainties reported by \textsc{SNID--SAGE}.

\begin{figure}
\centering
\includegraphics[width=\linewidth]{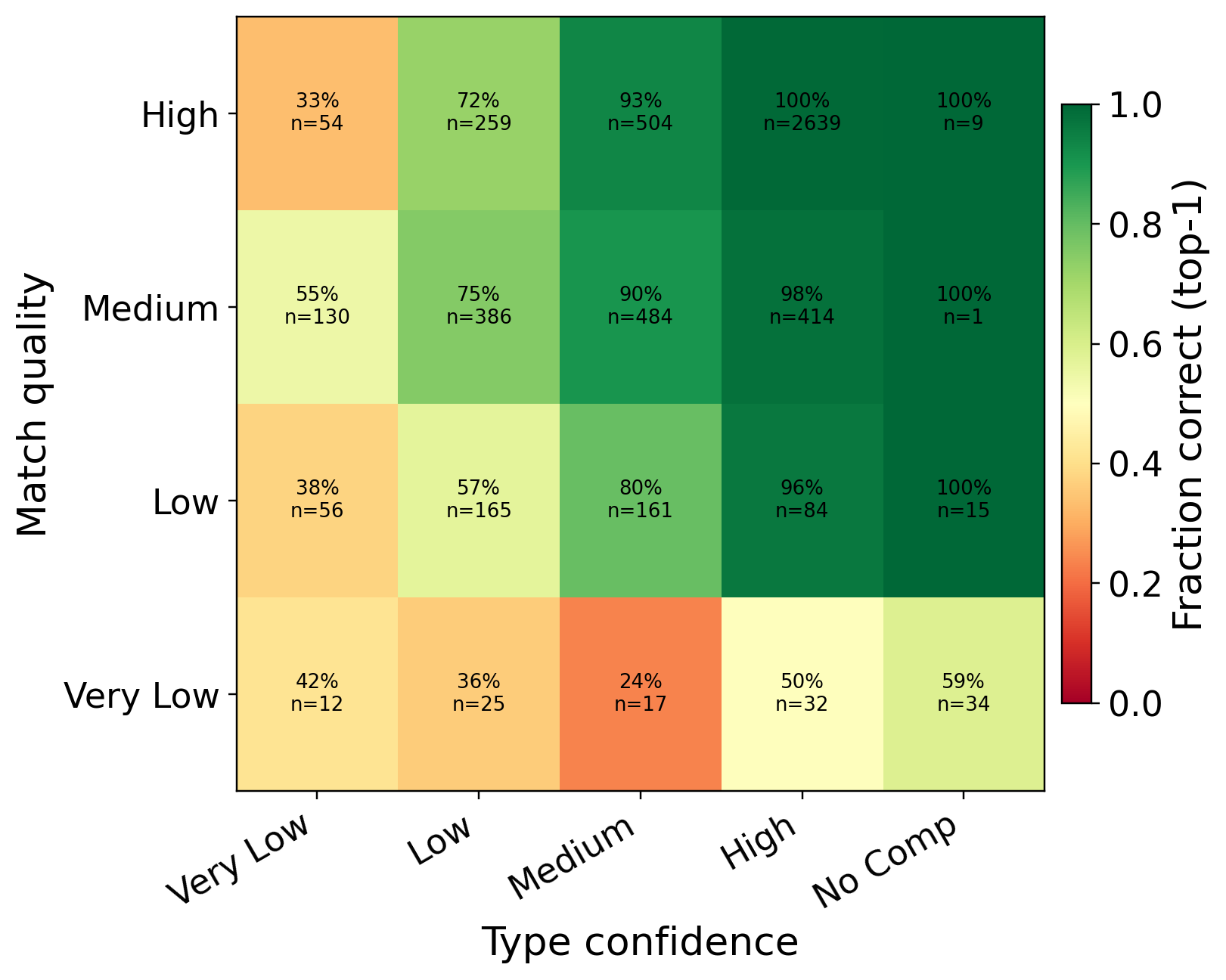}
\caption{\textbf{Type-recovery rate as a function of match quality and type confidence in the leave-one-out experiment.}
Each cell shows the fraction of spectra for which the correct type appears as the top-ranked final solution, together with the number of spectra contributing to that bin. Recovery improves strongly with increasing match quality, and the highest success rates occur when both the match quality and type-confidence class are high, indicating that the confidence metric provides a meaningful measure of classification reliability. The ``No Comp'' category denotes cases where no competing secondary solution is identified. This occurs when a single solution is clearly preferred, or when only one marginally plausible match is present and all other candidates are rejected as too low quality.}
\label{fig:loo_quality_confidence}
\end{figure}

To further assess the reliability of the classification outputs, we examine how the internally defined match-quality and type-confidence categories relate to actual classification performance in the leave-one-out experiment. Here, the type confidence quantifies the relative strength of the preferred solution compared to the next-best competing type, as introduced in Sec. \ref{sec:methods:gmm}. Figure~\ref{fig:loo_quality_confidence} shows the fraction of correctly recovered types as a function of these two quantities.

As expected, the classification reliability increases strongly when both the match-quality score and the type-confidence class are high. The best performance is obtained in the upper-right region of the plot, where the correct type is recovered for the overwhelming majority of spectra. By contrast, the \emph{Very Low} quality bins remain comparatively unreliable even when the confidence class is moderate or high, indicating that the quality score contains important information beyond the type-confidence flag alone. Taken together, these two quantities provide a compact and interpretable summary of classification reliability, enabling automated workflows to identify high-confidence results while flagging uncertain cases for further inspection.

Overall, the leave-one-out analysis demonstrates that \textsc{SNID--SAGE} provides robust and internally consistent estimates of type, redshift, and phase within the template library. Performance differences across transient classes are largely driven by template coverage: well-represented and spectrally homogeneous classes (e.g.\ Type~Ia) show the most stable behaviour, while rarer and more heterogeneous classes exhibit reduced performance. This suggests that further improvements can be achieved through the expansion and continued curation of the template library.

\subsection{WISeREP Database Analysis}
\label{sec:validation:wiserep}

To evaluate performance under realistic observational conditions, we analyse a large sample of astrophysical transient spectra from the WISeREP database \citep{yaron2012wiserep} with spectral coverage over the 4000--7000\,\AA\ interval, irrespective of spectral type. This sample comprises 46\,763 spectra spanning a wide range of object types, phases, and data quality, the vast majority of which are user-uploaded spectra.
This dataset serves two complementary purposes. First, it provides a large-scale public reference catalogue of \textsc{SNID--SAGE} classifications applied uniformly to a heterogeneous archival dataset. Second, it enables quantitative assessment of redshift recovery through comparison with independently determined host-galaxy redshifts where available.

The WISeREP database aggregates classifications from multiple surveys, instruments, and authors, and does not enforce a uniform taxonomy across all entries, therefore we do not perform a direct comparison between the predicted type/subtype labels and the WISeREP classifications. Instead, we provide the full \textsc{SNID--SAGE} set of classifications and derived quantities in a public catalogue \footnote{\url{https://fiorenst.github.io/SNID-SAGE/table/}} and leave detailed taxonomy comparisons to targeted user studies. The overall composition of inferred types and match-quality classes is summarised in Table~\ref{tab:wiserep_type_quality_nocuts}, providing a baseline view of the full WISeREP sample as processed by \textsc{SNID--SAGE}.

\begin{table}
    \centering
    \caption{Breakdown of main \textsc{SNID--SAGE} types and match-quality classes for the 45\,551 spectra that yielded valid \textsc{SNID--SAGE} redshift solutions from the WISeREP sample.}    
    \label{tab:wiserep_type_quality_nocuts}
    \begin{tabular}{lrrrr}
        \hline
        Main type & Very Low & Low & Medium & High \\
        \hline
        AGN & 17 & 42 & 77 & 92 \\
        GAP & 47 & 63 & 107 & 210 \\
        Galaxy & 83 & 265 & 351 & 388 \\
        II & 1\,071 & 2\,432 & 2\,992 & 4\,747 \\
        Ia & 918 & 4\,802 & 8\,864 & 10\,782 \\
        Ib & 137 & 306 & 472 & 596 \\
        Ibn & 64 & 75 & 123 & 154 \\
        Ic & 303 & 848 & 836 & 528 \\
        Icn & 23 & 5 & 7 & 19 \\
        KN & 13 & 2 & 0 & 3 \\
        LFBOT & 18 & 4 & 5 & 14 \\
        SLSN & 203 & 274 & 229 & 259 \\
        Star & 87 & 167 & 233 & 290 \\
        TDE & 141 & 170 & 230 & 363 \\
        \hline
    \end{tabular}
\end{table}

Redshift, by contrast, provides a well-defined quantity that can be validated against independent host-galaxy measurements, and we therefore focus on redshift recovery. Because transients are frequently offset from their hosts and may lie near multiple projected galaxies, secure host association is required before comparing inferred and catalogue redshifts.
To perform this association, we adopt host galaxies from the REGALADE catalogue \citep{Tranin2025}, a unified compilation of multiple galaxy surveys with uniformly derived redshift and galaxy-property measurements. Host association is carried out using the directional light radius (DLR) framework \citep{Gupta2016}, in which the projected transient–galaxy separation is normalised by the galaxy’s elliptical light profile. Candidate hosts are ranked by increasing DLR, with the minimum-DLR galaxy adopted as the reference host, and associations with $\mathrm{DLR} > 2.5$ are rejected. 

For this analysis, we restrict the sample to transients whose associated hosts have reliable spectroscopic redshifts. We further require consistency between the REGALADE and WISeREP redshift measurements, excluding objects with $|z_{\rm REGALADE} - z_{\rm WISeREP}| > 0.025$. This threshold is intentionally broader than the intrinsic redshift precision of the spectral pipeline and is designed to remove catastrophic host mismatches rather than small systematic offsets. Out of the initial 46\,763 spectra, 45\,551 yield a valid classification and redshift solution with \textsc{SNID--SAGE}. Of these, 25\,931 satisfy the host-association and redshift-consistency criteria. We additionally exclude WISeREP entries flagged as CAUTION and a small set of telescope/instrument combinations for which most spectra were uploaded already de-redshifted by users and therefore do not provide a meaningful test of redshift recovery. After all cuts, the final validation sample comprises 24\,223 spectra.

\begin{figure}
\centering
\includegraphics[width=\linewidth]{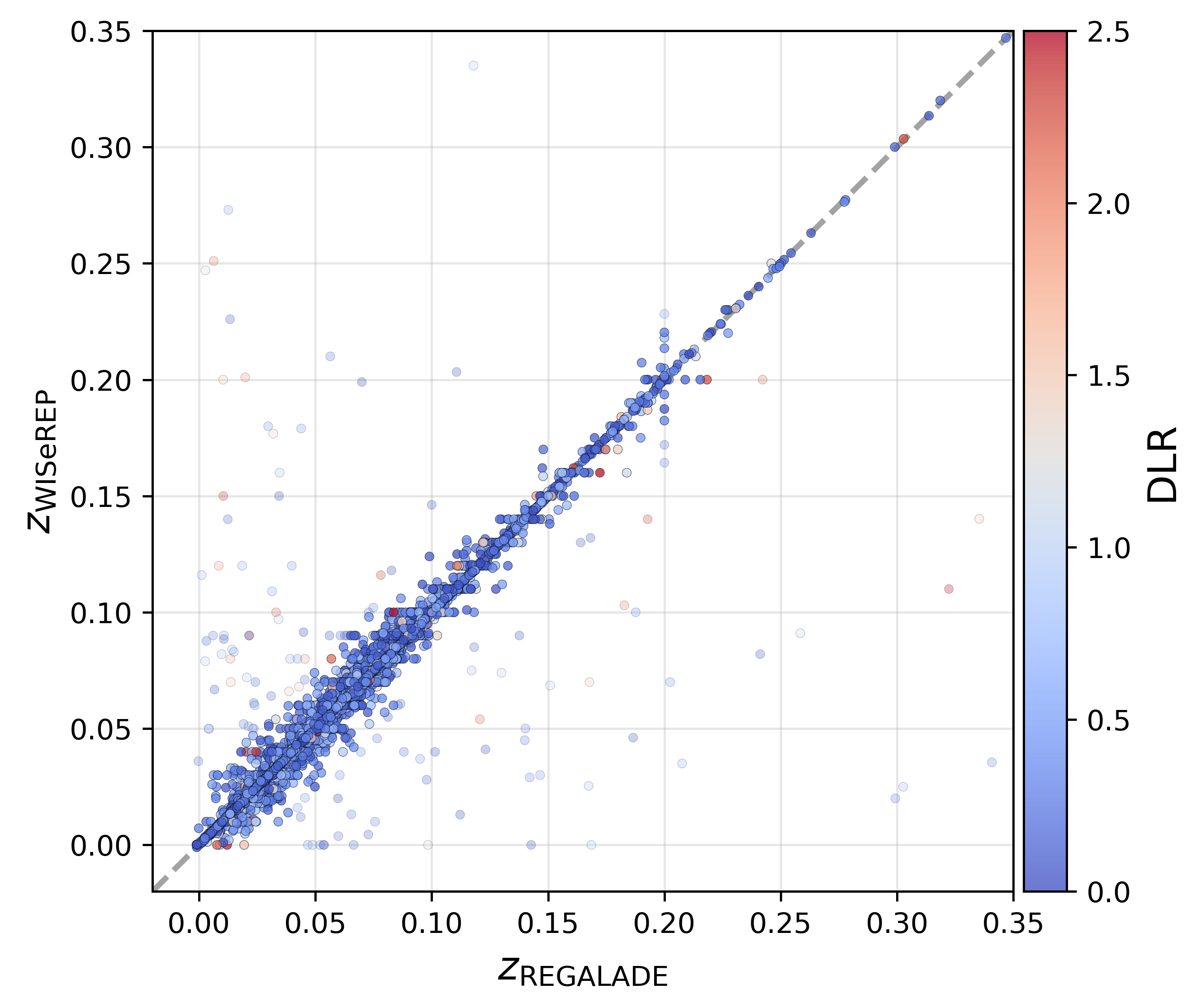}
\caption{
Comparison between WISeREP and REGALADE spectroscopic host-galaxy redshifts. Points are coloured by the host-association DLR. The dashed line shows the one-to-one relation. A small fraction of outliers reflects inconsistencies between catalogue redshifts and motivates the $|\Delta z_{\rm host}| < 0.025$ consistency cut used to define the final validation sample.
}
\label{fig:wiserep_regalade_host}
\end{figure}

\begin{figure*}
  \centering
  \includegraphics[width=\linewidth]{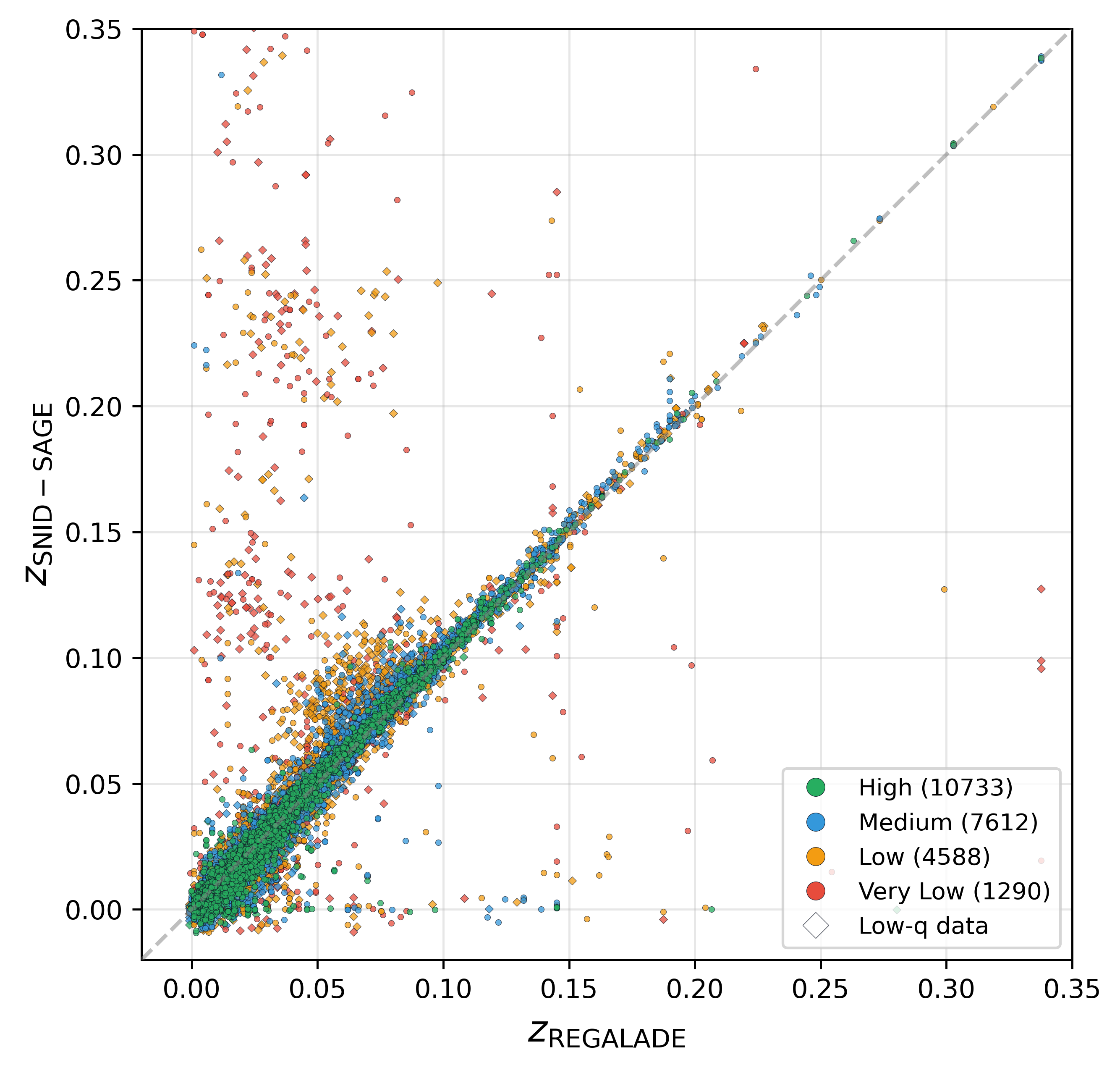}
  \caption{
    Comparison between \textsc{SNID--SAGE} redshifts and REGALADE host-galaxy reference redshifts for the full WISeREP validation sample. Points are colour-coded by \textsc{SNID--SAGE} match-quality class, and spectra from low-quality data instruments are shown as diamonds. The dashed line indicates the one-to-one relation. A small subset of spectra classified with \emph{High} match quality but with predicted redshifts near zero likely corresponds to spectra that have been pre-corrected to the rest frame prior to upload.
  }
  \label{fig:wiserep_zz_all}
\end{figure*}

Before comparing \textsc{SNID--SAGE} redshifts with host-galaxy measurements, we first evaluate the consistency between host redshifts reported in WISeREP and those extracted from REGALADE.  
Figure~\ref{fig:wiserep_regalade_host} shows the direct comparison between WISeREP-reported redshifts and the corresponding spectroscopic host redshifts from REGALADE for our sample. 
Although the overall agreement is good, a measurable intrinsic scatter is present. This is not unexpected: WISeREP redshifts are aggregated from multiple surveys and authors and are frequently user-reported, resulting in heterogeneous precision and occasional inconsistencies. The level of scatter between REGALADE and WISeREP is comparable to the residual dispersion measured later in this Section for \textsc{SNID--SAGE} redshift recovery (see Figure~\ref{fig:wiserep_zz_all}), indicating that part of the apparent redshift scatter in large-scale comparisons originates from catalogue-level differences rather than from the spectral inference itself.
Having defined the final validation sample (24\,223 spectra), we examine its composition in terms of inferred transient type and match quality. The resulting distribution is shown in Table~\ref{tab:wiserep_type_quality_val}.

\begin{table}
    \centering
    \caption{Breakdown of main \textsc{SNID--SAGE} types and match-quality restricted to the validation sample after all cuts (REGALADE matching, $DLR < 2.5$, $|\Delta z_{\mathrm{host}}| \le 0.025$, excluding WISeREP CAUTION spectra and specified problematic instruments). The total is 24\,223 spectra, as described in Section\,\ref{sec:validation:wiserep}.}
    \label{tab:wiserep_type_quality_val}
    \begin{tabular}{lrrrr}
        \hline
        Main type & Very Low & Low & Medium & High \\
        \hline
        AGN & 10 & 25 & 43 & 55 \\
        GAP & 36 & 47 & 53 & 120 \\
        Galaxy & 23 & 143 & 203 & 278 \\
        II & 466 & 1\,222 & 1\,476 & 2\,581 \\
        Ia & 392 & 2\,239 & 4\,636 & 6\,457 \\
        Ib & 63 & 199 & 339 & 408 \\
        Ibn & 26 & 41 & 46 & 94 \\
        Ic & 127 & 474 & 529 & 398 \\
        Icn & 11 & 5 & 6 & 18 \\
        KN & 5 & 0 & 0 & 0 \\
        LFBOT & 5 & 4 & 4 & 13 \\
        SLSN & 69 & 81 & 77 & 96 \\
        Star & 4 & 25 & 52 & 11 \\
        TDE & 53 & 83 & 148 & 204 \\
        \hline
    \end{tabular}
\end{table}

Relative to the raw WISeREP sample, the validation subset is mildly enriched in higher-quality matches, particularly for Types~Ia and II, reflecting the requirement of secure host associations and consistent catalogue redshifts. Nevertheless, the retained sample spans a broad range of transient classes, including stripped-envelope supernovae (Ib/Ic/Ibn), SLSN, TDEs, AGN, and other rarer populations. A substantial fraction of spectra in the major supernova categories achieve Medium or High match quality, indicating that the template library and clustering framework provide robust solutions across the dominant transient populations while maintaining coverage of rarer events.

Figure~\ref{fig:wiserep_zz_all} presents the core redshift validation result, comparing \textsc{SNID--SAGE} redshifts to REGALADE host-galaxy reference redshifts for the full validation sample of 24\,223 spectra. We also highlight a subset of spectra obtained with instruments that frequently yield lower-quality matches. These telescope/instrument combinations are defined empirically as those for which more than 40\% of spectra are assigned \emph{Very Low} or \emph{Low} match quality by \textsc{SNID--SAGE}. 
This subset is dominated by lower-resolution classification instruments, in particular Palomar P60/SEDM, which was designed for rapid, coarse classification and plays an important role in transient filtering despite its limited spectral resolution. The Liverpool Telescope SPRAT spectrograph and the UH88/SNIFS combination also fall into this category. Spectra from these instruments contribute disproportionately to the \emph{Very Low} and \emph{Low} match-quality regimes and account for a significant fraction of the outliers seen in Figure~\ref{fig:wiserep_zz_all}.

The majority of spectra cluster tightly around the one-to-one relation, with dispersion strongly correlated with match quality. High-quality matches exhibit minimal scatter and negligible systematic bias across the sampled redshift range. The Medium class remains well behaved, with only modest broadening, while the Low and Very Low classes show increasing dispersion and a higher incidence of catastrophic outliers. Spectra obtained with the instruments that we identify as producing lower quality data contribute disproportionately to the largest deviations. In particular, P60/SEDM displays a modest positive redshift bias that increases toward $z\simeq0.06$--0.08, consistent with operation near the practical sensitivity limit where spectral features are less constraining and cross-correlation solutions become more susceptible to template degeneracies.

Table~\ref{tab:wiserep_dzsage_quality_ref4567_allinst} quantifies the redshift residuals relative to REGALADE host-galaxy redshifts as a function of match quality. The median residual remains of order $10^{-3}$ across all match-quality classes, while the scatter decreases monotonically with increasing match quality. This behaviour confirms that the \textsc{SNID--SAGE} quality metric provides an effective empirical proxy for redshift reliability.

\begin{table}
    \centering
    \caption{Catalogue-level redshift agreement between \textsc{SNID--SAGE} redshifts and REGALADE host-galaxy redshifts for the WISeREP validation sample of 24\,223 spectra. We report $\Delta z_{\rm SAGE}\equiv z_{\rm SNID-SAGE}-z_{\rm REGALADE}$, giving the median and standard deviation within each \textsc{SNID--SAGE} match-quality class.}
    \label{tab:wiserep_dzsage_quality_ref4567_allinst}
    \begin{tabular}{lcc}
        \hline
        Match quality & Median $\Delta z_{\rm SAGE}$ & Std.\ $\Delta z_{\rm SAGE}$ \\
        \hline
        Very Low & 0.001613 & 0.130564 \\
        Low &  0.001490 & 0.056314 \\
        Medium & 0.001175 & 0.011160 \\
        High & 0.000563 & 0.005946 \\
        \hline
    \end{tabular}
\end{table}

Overall, the WISeREP validation demonstrates that, for spectra with reliable host associations and at least \emph{Medium} match quality, \textsc{SNID--SAGE} delivers robust and nearly unbiased redshift estimates across the sampled redshift range. At \emph{Very Low} match quality, catastrophic outliers are possible---typically arising from extremely low-information spectra (weak or absent features, strong contamination, or limited wavelength coverage), as well as from objects that are rare or sparsely represented in the template library. The \emph{Low} class remains usable in aggregate but carries a substantially higher outlier rate than \emph{Medium} and \emph{High}. 

The comparison between WISeREP and REGALADE host-galaxy redshifts (Figure~\ref{fig:wiserep_regalade_host}) shows that a non-negligible fraction of the observed residual scatter arises from catalogue-level inconsistencies, implying that part of the dispersion in \textsc{SNID--SAGE} redshift recovery is driven by external reference uncertainties rather than limitations of the spectral inference itself.

We note that the WISeREP validation sample is not fully independent of the template library, because a subset of spectra and/or objects represented in public archives may also contribute to the construction of the default template set. The leave-one-out analysis on the template library, therefore, provides the primary controlled validation of classification and phase performance, while the WISeREP analysis should be interpreted as a large-scale deployment and redshift-recovery test on a heterogeneous archival sample rather than as a strictly independent classification benchmark.

\section{Conclusions}
\label{sec:conclusions}

We have presented the Python-based \textsc{SNID--SAGE}, a framework for supernova spectral classification that builds on the established \textsc{SNID} cross-correlation methodology with practical and methodological refinements for survey-scale use. The pipeline combines reproducible preprocessing, an optimised cross-correlation engine, a new match-quality metric, and redshift-space clustering to consolidate multiple template matches into stable type and subtype assignments.

Validation through leave-one-out testing of the template library yields high overall type-recovery performance for the main supernova types, with near-perfect recovery for Type~Ia supernovae and stable redshift inference across the major classes represented in the library. Large-scale application to $\sim$46\,000 WISeREP spectra, including a conservative validation subset of 24\,223 spectra with secure host associations, demonstrates robust and nearly unbiased redshift recovery over the sampled redshift range.

Performance remains dependent on template coverage and spectral diversity. Recovery rates are reduced for intrinsically heterogeneous or sparsely represented classes, and very low-information spectra can produce catastrophic outliers at \emph{Very Low} match quality. Continued expansion and curation of the template library will therefore be essential, particularly for rare or emerging transient populations and for improving performance in low signal-to-noise regimes.

\textsc{SNID--SAGE} is publicly available with full documentation and installation instructions, supporting both interactive analysis and scalable deployment in spectroscopic survey workflows.

\section*{Acknowledgements}

F.S. acknowledges support from the Royal Society Newton International Fellowship NIFR1241769.  
SJS acknowledges funding from STFC Grants ST/Y001605/1, ST/X006506/1, ST/T000198/1, a Royal Society Research Professorship and the Hintze Charitable Foundation.
We thank Stephane Blondin for helpful discussions, and Hugo Tranin for assistance with the REGALADE catalogue.

\section*{Data Availability}

The \textsc{SNID--SAGE} source code is publicly available at \url{https://github.com/FiorenSt/SNID-SAGE}, together with installation instructions and full user documentation. 

The WISeREP classification catalogue produced in this work, including inferred types, redshifts, phases, match-quality metrics, and confidence estimates, is available in interactive form at \url{https://fiorenst.github.io/SNID-SAGE/table/}. 

The external datasets analysed in this study (WISeREP spectra and REGALADE host-galaxy redshifts) are publicly accessible through their respective archives, subject to their data access policies.



\bibliographystyle{mnras}
\bibliography{snid_sage_refs}

@ARTICLE{tuckerSCAT2022,
       author = {{Tucker}, M.~A. and {Shappee}, B.~J. and {Huber}, M.~E. and {Payne}, A.~V. and {Do}, A. and {Hinkle}, J.~T. and {de Jaeger}, T. and {Ashall}, C. and {Desai}, D.~D. and {Hoogendam}, W.~B. and {Aldering}, G. and {Auchettl}, K. and {Baranec}, C. and {Bulger}, J. and {Chambers}, K. and {Chun}, M. and {Hodapp}, K.~W. and {Lowe}, T.~B. and {McKay}, L. and {Rampy}, R. and {Rubin}, D. and {Tonry}, J.~L.},
        title = "{The Spectroscopic Classification of Astronomical Transients (SCAT) Survey: Overview, Pipeline Description, Initial Results, and Future Plans}",
      journal = {\pasp},
         year = 2022,
        month = dec,
       volume = {134},
       number = {1042},
          eid = {124502},
        pages = {124502},
          doi = {10.1088/1538-3873/aca719},
archivePrefix = {arXiv},
       eprint = {2210.09322},
 primaryClass = {astro-ph.IM},
       adsurl = {https://ui.adsabs.harvard.edu/abs/2022PASP..134l4502T}
}

@ARTICLE{cepaOSIRIS1998,
       author = {{Cepa}, Jordi},
        title = "{OSIRIS Imaging and Spectroscopy for the GTC}",
      journal = {\apss},
         year = 1998,
        month = jun,
       volume = {263},
        pages = {369-372},
          doi = {10.1023/A:1002144913887},
       adsurl = {https://ui.adsabs.harvard.edu/abs/1998Ap&SS.263..369C}
}

@ARTICLE{appenzellerFORS1998,
       author = {{Appenzeller}, I. and {Fricke}, K. and {F{\"u}rtig}, W. and {G{\"a}ssler}, W. and {H{\"a}fner}, R. and {Harke}, R. and {Hess}, H.-J. and {Hummel}, W. and {J{\"u}rgens}, P. and {Kudritzki}, R.-P. and {Mantel}, K.-H. and {Meisl}, W. and {Muschielok}, B. and {Nicklas}, H. and {Rupprecht}, G. and {Seifert}, W. and {Stahl}, O. and {Szeifert}, T. and {Tarantik}, K.},
        title = "{Successful commissioning of FORS1 - the first optical instrument on the VLT.}",
      journal = {The Messenger},
         year = 1998,
        month = dec,
       volume = {94},
        pages = {1-6},
       adsurl = {https://ui.adsabs.harvard.edu/abs/1998Msngr..94....1A}
}

@ARTICLE{okeLRIS1995,
       author = {{Oke}, J.~B. and {Cohen}, J.~G. and {Carr}, M. and {Cromer}, J. and {Dingizian}, A. and {Harris}, F.~H. and {Labrecque}, S. and {Lucinio}, R. and {Schaal}, W. and {Epps}, H. and {Miller}, J.},
        title = "{The Keck Low-Resolution Imaging Spectrometer}",
      journal = {\pasp},
         year = 1995,
        month = apr,
       volume = {107},
        pages = {375},
          doi = {10.1086/133562},
       adsurl = {https://ui.adsabs.harvard.edu/abs/1995PASP..107..375O}
}

@article{tonry2018atlas,
  author       = {Tonry, J. L. and Denneau, L. and Heinze, A. N. and Stalder, B. and Smith, K. W. and Smartt, S. J. and Stubbs, C. W. and Weiland, H. J. and Rest, A.},
  title        = {ATLAS: A High-cadence All-sky Survey System},
  journal      = {Publications of the Astronomical Society of the Pacific},
  year         = {2018},
  volume       = {130},
  number       = {988},
  eid          = {064505},
  doi          = {10.1088/1538-3873/aabadf},
  url          = {https://iopscience.iop.org/article/10.1088/1538-3873/aabadf},
  adsurl       = {https://ui.adsabs.harvard.edu/abs/2018PASP..130f4505T},
  bibcode      = {2018PASP..130f4505T}
}

@article{groot2024blackgem,
  author       = {Groot, P. J. and Bloemen, S. and Vreeswijk, P. M. and van Roestel, J. C. J. and Jonker, P. G. and Nelemans, G. and et al.},
  title        = {The BlackGEM Telescope Array. I. Overview},
  journal      = {Publications of the Astronomical Society of the Pacific},
  year         = {2024},
  volume       = {136},
  number       = {11},
  eid          = {115003},
  doi          = {10.1088/1538-3873/ad8b6a}
}

@article{steeghs2022goto,
  author       = {Steeghs, D. and Galloway, D. K. and Ackley, K. and et al.},
  title        = {The Gravitational-wave Optical Transient Observer (GOTO): Prototype performance and prospects for transient science},
  journal      = {Monthly Notices of the Royal Astronomical Society},
  year         = {2022},
  volume       = {511},
  number       = {2},
  pages        = {2405--2422},
  doi          = {10.1093/mnras/stac013},
  url          = {https://ui.adsabs.harvard.edu/abs/2022MNRAS.511.2405S/abstract},
  adsurl       = {https://ui.adsabs.harvard.edu/abs/2022MNRAS.511.2405S},
  bibcode      = {2022MNRAS.511.2405S}
}

@article{flewelling2020ps1,
  author       = {Flewelling, H. A. and Magnier, E. A. and Chambers, K. C. and Heasley, J. N. and Holmberg, C. and Huber, M. E. and et al.},
  title        = {The Pan-STARRS1 Database and Data Products},
  journal      = {The Astrophysical Journal Supplement Series},
  year         = {2020},
  volume       = {251},
  number       = {1},
  eid          = {7},
  doi          = {10.3847/1538-4365/abb82d},
  url          = {https://iopscience.iop.org/article/10.3847/1538-4365/abb82d},
  adsurl       = {https://ui.adsabs.harvard.edu/abs/2020ApJS..251....7F},
  bibcode      = {2020ApJS..251....7F}
}

@article{bellm2019ztf,
  author       = {Bellm, E. C. and Kulkarni, S. R. and Graham, M. J. and Dekany, R. and Smith, R. M. and Riddle, R. and et al.},
  title        = {The Zwicky Transient Facility: System Overview, Performance, and First Results},
  journal      = {Publications of the Astronomical Society of the Pacific},
  year         = {2019},
  volume       = {131},
  number       = {1001},
  eid          = {018002},
  doi          = {10.1088/1538-3873/aaecbe},
  url          = {https://ui.adsabs.harvard.edu/abs/2019PASP..131a8002B/abstract},
  adsurl       = {https://ui.adsabs.harvard.edu/abs/2019PASP..131a8002B},
  bibcode      = {2019PASP..131a8002B}
}

@article{ivezic2019lsst,
  author       = {Ivezi{\'c}, {\v Z}. and Kahn, S. M. and Tyson, J. A. and et al.},
  title        = {LSST: From Science Drivers to Reference Design and Anticipated Data Products},
  journal      = {The Astrophysical Journal},
  year         = {2019},
  volume       = {873},
  number       = {2},
  eid          = {111},
  doi          = {10.3847/1538-4357/ab042c},
  url          = {https://iopscience.iop.org/article/10.3847/1538-4357/ab042c},
  adsurl       = {https://ui.adsabs.harvard.edu/abs/2019ApJ...873..111I},
  bibcode      = {2019ApJ...873..111I}
}

@article{smartt2015pessto,
  author       = {Smartt, S. J. and Valenti, S. and Fraser, M. and et al.},
  title        = {PESSTO: Public ESO Spectroscopic Survey of Transient Objects},
  journal      = {Astronomy \& Astrophysics},
  year         = {2015},
  volume       = {579},
  eid          = {A40},
  doi          = {10.1051/0004-6361/201423884},
  adsurl       = {https://ui.adsabs.harvard.edu/abs/2015A&A...579A..40S},
  url          = {https://www.aanda.org/articles/aa/abs/2015/07/aa23884-14/aa23884-14.html},
  bibcode      = {2015A&A...579A..40S}
}

@article{blondin2007snid,
  author       = {Blondin, S. and Tonry, J. L.},
  title        = {Determining the Type, Redshift, and Age of a Supernova Spectrum},
  journal      = {The Astrophysical Journal},
  year         = {2007},
  volume       = {666},
  number       = {2},
  pages        = {1024--1047},
  doi          = {10.1086/520494},
  url          = {https://iopscience.iop.org/issue/0004-637X/666/2},
  adsurl       = {https://ui.adsabs.harvard.edu/abs/2007ApJ...666.1024B},
  bibcode      = {2007ApJ...666.1024B}
}

@article{howell2005superfit,
  author       = {Howell, D. A. and Sullivan, M. and Perrett, K. and Bronder, T. J. and Hook, I. M. and Astier, P. and et al.},
  title        = {Gemini Spectroscopy of Supernovae from the Supernova Legacy Survey: Improving High-Redshift Supernova Selection and Classification},
  journal      = {The Astrophysical Journal},
  year         = {2005},
  volume       = {634},
  number       = {2},
  pages        = {1190--1201},
  doi          = {10.1086/497119},
  url          = {https://iopscience.iop.org/article/10.1086/497119},
  adsurl       = {https://ui.adsabs.harvard.edu/abs/2005ApJ...634.1190H/abstract},
  bibcode      = {2005ApJ...634.1190H}
}

@article{harutyunyan2008gelato,
  author       = {Harutyunyan, A. H. and Pfahler, P. and Pastorello, A. and Taubenberger, S. and Turatto, M. and Cappellaro, E. and Benetti, S. and Elias-Rosa, N. and Navasardyan, H. and Valenti, S. and et al.},
  title        = {ESC supernova spectroscopy of non-ESC targets (GELATO)},
  journal      = {Astronomy \& Astrophysics},
  year         = {2008},
  volume       = {488},
  number       = {1},
  pages        = {383--399},
  doi          = {10.1051/0004-6361:20078859},
  url          = {https://www.aanda.org/articles/aa/abs/2008/34/aa8859-07/aa8859-07.html},
  adsurl       = {https://ui.adsabs.harvard.edu/abs/2008A&A...488..383H/abstract},
  bibcode      = {2008A&A...488..383H}
}

@article{muthukrishna2019dash,
  author       = {Muthukrishna, D. and Parkinson, D. and Tucker, B. E.},
  title        = {DASH: Deep Learning for the Automated Spectral Classification of Supernovae and Their Hosts},
  journal      = {The Astrophysical Journal},
  year         = {2019},
  volume       = {885},
  eid          = {85},
  doi          = {10.3847/1538-4357/ab48f4},
  url          = {https://iopscience.iop.org/article/10.3847/1538-4357/ab48f4},
  adsurl       = {https://ui.adsabs.harvard.edu/abs/2019ApJ...885...85M/abstract},
  bibcode      = {2019ApJ...885...85M}
}

@article{yaron2012wiserep,
  author       = {Yaron, O. and Gal-Yam, A.},
  title        = {WISeREP---An Interactive Supernova Data Repository},
  journal      = {Publications of the Astronomical Society of the Pacific},
  year         = {2012},
  volume       = {124},
  number       = {917},
  pages        = {668--681},
  doi          = {10.1086/666656},
  url          = {https://iopscience.iop.org/article/10.1086/666656},
  adsurl       = {https://ui.adsabs.harvard.edu/abs/2012PASP..124..668Y/abstract},
  bibcode      = {2012PASP..124..668Y}
}

@misc{goldwasser2022ngsf,
  author       = {Goldwasser, S. and Bianco, F. and Sharma, Y. and Goobar, A. and Magee, M. and Sollerman, J. and others},
  title        = {NGSF: A New Generation of Supernova Fitter},
  howpublished = {Transient Name Server AstroNote, No.\ 191},
  year         = {2022},
  url          = {https://www.wis-tns.org/astronotes/astronote/191},
  adsurl       = {https://ui.adsabs.harvard.edu/abs/2022TNSAN.191....1G/abstract},
  bibcode      = {2022TNSAN.191....1G}
}

@article{tonry1979,
  author   = {Tonry, J. and Davis, M.},
  title    = {A Survey of Galaxy Redshifts. I. Data Reduction Techniques},
  journal  = {The Astronomical Journal},
  year     = {1979},
  volume   = {84},
  pages    = {1511--1525},
  doi      = {10.1086/112569},
  url      = {https://doi.org/10.1086/112569},
  adsurl   = {https://ui.adsabs.harvard.edu/abs/1979AJ.....84.1511T/abstract},
  bibcode  = {1979AJ.....84.1511T}
}

@article{harris1978,
  author   = {Harris, Fredric J.},
  title    = {On the Use of Windows for Harmonic Analysis with the Discrete Fourier Transform},
  journal  = {Proceedings of the IEEE},
  year     = {1978},
  volume   = {66},
  number   = {1},
  pages    = {51--83},
  doi      = {10.1109/PROC.1978.10837},
  url      = {https://doi.org/10.1109/PROC.1978.10837},
  adsurl   = {https://ui.adsabs.harvard.edu/abs/1978IEEEP..66...51H/abstract},
  bibcode  = {1978IEEEP..66...51H}
}

@article{savitzky1964,
  author   = {Savitzky, Abraham and Golay, M. J. E.},
  title    = {Smoothing and Differentiation of Data by Simplified Least Squares Procedures},
  journal  = {Analytical Chemistry},
  year     = {1964},
  volume   = {36},
  number   = {8},
  pages    = {1627--1639},
  doi      = {10.1021/ac60214a047},
  url      = {https://doi.org/10.1021/ac60214a047},
  adsurl   = {https://ui.adsabs.harvard.edu/abs/1964AnaCh..36.1627S/abstract},
  bibcode  = {1964AnaCh..36.1627S}
}

@article{lin1989,
  author   = {Lin, Lawrence I.-K.},
  title    = {A Concordance Correlation Coefficient to Evaluate Reproducibility},
  journal  = {Biometrics},
  year     = {1989},
  volume   = {45},
  number   = {1},
  pages    = {255--268},
  doi      = {10.2307/2532051},
  url      = {https://doi.org/10.2307/2532051}
}

@article{schwarz1978bic,
  author   = {Schwarz, Gideon},
  title    = {Estimating the Dimension of a Model},
  journal  = {The Annals of Statistics},
  year     = {1978},
  volume   = {6},
  number   = {2},
  pages    = {461--464},
  doi      = {10.1214/aos/1176344136},
  url      = {https://doi.org/10.1214/aos/1176344136},
  adsurl   = {https://ui.adsabs.harvard.edu/abs/1978AnSta...6..461S/abstract},
  bibcode  = {1978AnSta...6..461S}
}

@ARTICLE{Shoko2024,
       author = {{Jin}, Shoko and {Trager}, Scott C. and {Dalton}, Gavin B. and {Aguerri}, J. Alfonso L. and {Drew}, J.~E. and {Falc{\'o}n-Barroso}, Jes{\'u}s and {G{\"a}nsicke}, Boris T. and {Hill}, Vanessa and {Iovino}, Angela and {Pieri}, Matthew M. and {Poggianti}, Bianca M. and {Smith}, D.~J.~B. and {Vallenari}, Antonella and {Abrams}, Don Carlos and {Aguado}, David S. and {Antoja}, Teresa and {Arag{\'o}n-Salamanca}, Alfonso and {Ascasibar}, Yago and {Babusiaux}, Carine and {Balcells}, Marc and {Barrena}, R. and {Battaglia}, Giuseppina and {Belokurov}, Vasily and {Bensby}, Thomas and {Bonifacio}, Piercarlo and {Bragaglia}, Angela and {Carrasco}, Esperanza and {Carrera}, Ricardo and {Cornwell}, Daniel J. and {Dom{\'\i}nguez-Palmero}, Lilian and {Duncan}, Kenneth J. and {Famaey}, Benoit and {Fari{\~n}a}, Cecilia and {Gonzalez}, Oscar A. and {Guest}, Steve and {Hatch}, Nina A. and {Hess}, Kelley M. and {Hoskin}, Matthew J. and {Irwin}, Mike and {Knapen}, Johan H. and {Koposov}, Sergey E. and {Kuchner}, Ulrike and {Laigle}, Clotilde and {Lewis}, Jim and {Longhetti}, Marcella and {Lucatello}, Sara and {M{\'e}ndez-Abreu}, Jairo and {Mercurio}, Amata and {Molaeinezhad}, Alireza and {Mongui{\'o}}, Maria and {Morrison}, Sean and {Murphy}, David N.~A. and {Peralta de Arriba}, Luis and {P{\'e}rez}, Isabel and {P{\'e}rez-R{\`a}fols}, Ignasi and {Pic{\'o}}, Sergio and {Raddi}, Roberto and {Romero-G{\'o}mez}, Merc{\`e} and {Royer}, Fr{\'e}d{\'e}ric and {Siebert}, Arnaud and {Seabroke}, George M. and {Som}, Debopam and {Terrett}, David and {Thomas}, Guillaume and {Wesson}, Roger and {Worley}, C. Clare and {Alfaro}, Emilio J. and {Allende Prieto}, Carlos and {Alonso-Santiago}, Javier and {Amos}, Nicholas J. and {Ashley}, Richard P. and {Balaguer-N{\'u}{\~n}ez}, Lola and {Balbinot}, Eduardo and {Bellazzini}, Michele and {Benn}, Chris R. and {Berlanas}, Sara R. and {Bernard}, Edouard J. and {Best}, Philip and {Bettoni}, Daniela and {Bianco}, Andrea and {Bishop}, Georgia and {Blomqvist}, Michael and {Boeche}, Corrado and {Bolzonella}, Micol and {Bonoli}, Silvia and {Bosma}, Albert and {Britavskiy}, Nikolay and {Busarello}, Gianni and {Caffau}, Elisabetta and {Cantat-Gaudin}, Tristan and {Castro-Ginard}, Alfred and {Couto}, Guilherme and {Carbajo-Hijarrubia}, Juan and {Carter}, David and {Casamiquela}, Laia and {Conrado}, Ana M. and {Corcho-Caballero}, Pablo and {Costantin}, Luca and {Deason}, Alis and {de Burgos}, Abel and {De Grandi}, Sabrina and {Di Matteo}, Paola and {Dom{\'\i}nguez-G{\'o}mez}, Jes{\'u}s and {Dorda}, Ricardo and {Drake}, Alyssa and {Dutta}, Rajeshwari and {Erkal}, Denis and {Feltzing}, Sofia and {Ferr{\'e}-Mateu}, Anna and {Feuillet}, Diane and {Figueras}, Francesca and {Fossati}, Matteo and {Franciosini}, Elena and {Frasca}, Antonio and {Fumagalli}, Michele and {Gallazzi}, Anna and {Garc{\'\i}a-Benito}, Rub{\'e}n and {Gentile Fusillo}, Nicola and {Gebran}, Marwan and {Gilbert}, James and {Gledhill}, T.~M. and {Gonz{\'a}lez Delgado}, Rosa M. and {Greimel}, Robert and {Guarcello}, Mario Giuseppe and {Guerra}, Jose and {Gullieuszik}, Marco and {Haines}, Christopher P. and {Hardcastle}, Martin J. and {Harris}, Amy and {Haywood}, Misha and {Helmi}, Amina and {Hernandez}, Nauzet and {Herrero}, Artemio and {Hughes}, Sarah and {Ir{\v{s}}i{\v{c}}}, Vid and {Jablonka}, Pascale and {Jarvis}, Matt J. and {Jordi}, Carme and {Kondapally}, Rohit and {Kordopatis}, Georges and {Krogager}, Jens-Kristian and {La Barbera}, Francesco and {Lam}, Man I. and {Larsen}, S{\o}ren S. and {Lemasle}, Bertrand and {Lewis}, Ian J. and {Lhom{\'e}}, Emilie and {Lind}, Karin and {Lodi}, Marcello and {Longobardi}, Alessia and {Lonoce}, Ilaria and {Magrini}, Laura and {Ma{\'\i}z Apell{\'a}niz}, Jes{\'u}s and {Marchal}, Olivier and {Marco}, Amparo and {Martin}, Nicolas F. and {Matsuno}, Tadafumi and {Maurogordato}, Sophie and {Merluzzi}, Paola and {Miralda-Escud{\'e}}, Jordi and {Molinari}, Emilio and {Monari}, Giacomo and {Morelli}, Lorenzo and {Mottram}, Christopher J. and {Naylor}, Tim and {Negueruela}, Ignacio and {O{\~n}orbe}, Jose and {Pancino}, Elena and {Peirani}, S{\'e}bastien and {Peletier}, Reynier F. and {Pozzetti}, Lucia and {Rainer}, Monica and {Ramos}, Pau and {Read}, Shaun C. and {Rossi}, Elena Maria and {R{\"o}ttgering}, Huub J.~A. and {Rubi{\~n}o-Mart{\'\i}n}, Jose Alberto and {Sabater}, Jose and {San Juan}, Jos{\'e} and {Sanna}, Nicoletta and {Schallig}, Ellen and {Schiavon}, Ricardo P. and {Schultheis}, Mathias and {Serra}, Paolo and {Shimwell}, Timothy W. and {Sim{\'o}n-D{\'\i}az}, Sergio and {Smith}, Russell J. and {Sordo}, Rosanna and {Sorini}, Daniele and {Soubiran}, Caroline and {Starkenburg}, Else and {Steele}, Iain A. and {Stott}, John and {Stuik}, Remko and {Tolstoy}, Eline and {Tortora}, Crescenzo and {Tsantaki}, Maria and {Van der Swaelmen}, Mathieu and {van Weeren}, Reinout J. and {Vergani}, Daniela},
        title = "{The wide-field, multiplexed, spectroscopic facility WEAVE: Survey design, overview, and simulated implementation}",
      journal = {\mnras},
         year = 2024,
        month = may,
       volume = {530},
       number = {3},
        pages = {2688-2730},
          doi = {10.1093/mnras/stad557},
archivePrefix = {arXiv},
       eprint = {2212.03981},
 primaryClass = {astro-ph.IM},
       adsurl = {https://ui.adsabs.harvard.edu/abs/2024MNRAS.530.2688J}
}

@ARTICLE{Frohmaier2025,
       author = {{Frohmaier}, C. and {Vincenzi}, M. and {Sullivan}, M. and {H{\"o}nig}, S.~F. and {Smith}, M. and {Addison}, H. and {Collett}, T. and {Dimitriadis}, G. and {Ellis}, R.~S. and {Gandhi}, P. and {Graur}, O. and {Hook}, I. and {Kelsey}, L. and {Kim}, Y.-L. and {Lidman}, C. and {Maguire}, K. and {Makrygianni}, L. and {Martin}, B. and {M{\"o}ller}, A. and {Nichol}, R.~C. and {Nicholl}, M. and {Schady}, P. and {Simmons}, B.~D. and {Smartt}, S.~J. and {Tempel}, E. and {Wiseman}, P. and {the LSST Dark Energy Science Collaboration}},
        title = "{TiDES: The 4MOST Time Domain Extragalactic Survey}",
      journal = {\apj},
         year = 2025,
        month = oct,
       volume = {992},
       number = {1},
          eid = {158},
        pages = {158},
          doi = {10.3847/1538-4357/adff4e},
archivePrefix = {arXiv},
       eprint = {2501.16311},
 primaryClass = {astro-ph.HE},
       adsurl = {https://ui.adsabs.harvard.edu/abs/2025ApJ...992..158F}
}

@ARTICLE{Dejong2019,
       author = {{de Jong}, R.~S. and {Agertz}, O. and {Berbel}, A.~A. and {Aird}, J. and {Alexander}, D.~A. and {Amarsi}, A. and {Anders}, F. and {Andrae}, R. and {Ansarinejad}, B. and {Ansorge}, W. and {Antilogus}, P. and {Anwand-Heerwart}, H. and {Arentsen}, A. and {Arnadottir}, A. and {Asplund}, M. and {Auger}, M. and {Azais}, N. and {Baade}, D. and {Baker}, G. and {Baker}, S. and {Balbinot}, E. and {Baldry}, I.~K. and {Banerji}, M. and {Barden}, S. and {Barklem}, P. and {Barth{\'e}l{\'e}my-Mazot}, E. and {Battistini}, C. and {Bauer}, S. and {Bell}, C.~P.~M. and {Bellido-Tirado}, O. and {Bellstedt}, S. and {Belokurov}, V. and {Bensby}, T. and {Bergemann}, M. and {Bestenlehner}, J.~M. and {Bielby}, R. and {Bilicki}, M. and {Blake}, C. and {Bland-Hawthorn}, J. and {Boeche}, C. and {Boland}, W. and {Boller}, T. and {Bongard}, S. and {Bongiorno}, A. and {Bonifacio}, P. and {Boudon}, D. and {Brooks}, D. and {Brown}, M.~J.~I. and {Brown}, R. and {Br{\"u}ggen}, M. and {Brynnel}, J. and {Brzeski}, J. and {Buchert}, T. and {Buschkamp}, P. and {Caffau}, E. and {Caillier}, P. and {Carrick}, J. and {Casagrande}, L. and {Case}, S. and {Casey}, A. and {Cesarini}, I. and {Cescutti}, G. and {Chapuis}, D. and {Chiappini}, C. and {Childress}, M. and {Christlieb}, N. and {Church}, R. and {Cioni}, M.-R.~L. and {Cluver}, M. and {Colless}, M. and {Collett}, T. and {Comparat}, J. and {Cooper}, A. and {Couch}, W. and {Courbin}, F. and {Croom}, S. and {Croton}, D. and {Daguis{\'e}}, E. and {Dalton}, G. and {Davies}, L.~J.~M. and {Davis}, T. and {de Laverny}, P. and {Deason}, A. and {Dionies}, F. and {Disseau}, K. and {Doel}, P. and {D{\"o}scher}, D. and {Driver}, S.~P. and {Dwelly}, T. and {Eckert}, D. and {Edge}, A. and {Edvardsson}, B. and {Youssoufi}, D.~E. and {Elhaddad}, A. and {Enke}, H. and {Erfanianfar}, G. and {Farrell}, T. and {Fechner}, T. and {Feiz}, C. and {Feltzing}, S. and {Ferreras}, I. and {Feuerstein}, D. and {Feuillet}, D. and {Finoguenov}, A. and {Ford}, D. and {Fotopoulou}, S. and {Fouesneau}, M. and {Frenk}, C. and {Frey}, S. and {Gaessler}, W. and {Geier}, S. and {Gentile Fusillo}, N. and {Gerhard}, O. and {Giannantonio}, T. and {Giannone}, D. and {Gibson}, B. and {Gillingham}, P. and {Gonz{\'a}lez-Fern{\'a}ndez}, C. and {Gonzalez-Solares}, E. and {Gottloeber}, S. and {Gould}, A. and {Grebel}, E.~K. and {Gueguen}, A. and {Guiglion}, G. and {Haehnelt}, M. and {Hahn}, T. and {Hansen}, C.~J. and {Hartman}, H. and {Hauptner}, K. and {Hawkins}, K. and {Haynes}, D. and {Haynes}, R. and {Heiter}, U. and {Helmi}, A. and {Aguayo}, C.~H. and {Hewett}, P. and {Hinton}, S. and {Hobbs}, D. and {Hoenig}, S. and {Hofman}, D. and {Hook}, I. and {Hopgood}, J. and {Hopkins}, A. and {Hourihane}, A. and {Howes}, L. and {Howlett}, C. and {Huet}, T. and {Irwin}, M. and {Iwert}, O. and {Jablonka}, P. and {Jahn}, T. and {Jahnke}, K. and {Jarno}, A. and {Jin}, S. and {Jofre}, P. and {Johl}, D. and {Jones}, D. and {J{\"o}nsson}, H. and {Jordan}, C. and {Karovicova}, I. and {Khalatyan}, A. and {Kelz}, A. and {Kennicutt}, R. and {King}, D. and {Kitaura}, F. and {Klar}, J. and {Klauser}, U. and {Kneib}, J.-P. and {Koch}, A. and {Koposov}, S. and {Kordopatis}, G. and {Korn}, A. and {Kosmalski}, J. and {Kotak}, R. and {Kovalev}, M. and {Kreckel}, K. and {Kripak}, Y. and {Krumpe}, M. and {Kuijken}, K. and {Kunder}, A. and {Kushniruk}, I. and {Lam}, M.~I. and {Lamer}, G. and {Laurent}, F. and {Lawrence}, J. and {Lehmitz}, M. and {Lemasle}, B. and {Lewis}, J. and {Li}, B. and {Lidman}, C. and {Lind}, K. and {Liske}, J. and {Lizon}, J.-L. and {Loveday}, J. and {Ludwig}, H.-G. and {McDermid}, R.~M. and {Maguire}, K. and {Mainieri}, V. and {Mali}, S. and {Mandel}, H.},
        title = "{4MOST: Project overview and information for the First Call for Proposals}",
      journal = {The Messenger},
         year = 2019,
        month = mar,
       volume = {175},
        pages = {3-11},
          doi = {10.18727/0722-6691/5117},
archivePrefix = {arXiv},
       eprint = {1903.02464},
 primaryClass = {astro-ph.IM},
       adsurl = {https://ui.adsabs.harvard.edu/abs/2019Msngr.175....3D}
}

@INPROCEEDINGS{Tamura2016,
       author = {{Tamura}, Naoyuki and {Takato}, Naruhisa and {Shimono}, Atsushi and {Moritani}, Yuki and {Yabe}, Kiyoto and {Ishizuka}, Yuki and {Ueda}, Akitoshi and {Kamata}, Yukiko and {Aghazarian}, Hrand and {Arnouts}, St{\'e}phane and {Barban}, Gabriel and {Barkhouser}, Robert H. and {Borges}, Renato C. and {Braun}, David F. and {Carr}, Michael A. and {Chabaud}, Pierre-Yves and {Chang}, Yin-Chang and {Chen}, Hsin-Yo and {Chiba}, Masashi and {Chou}, Richard C.~Y. and {Chu}, You-Hua and {Cohen}, Judith and {de Almeida}, Rodrigo P. and {de Oliveira}, Antonio C. and {de Oliveira}, Ligia S. and {Dekany}, Richard G. and {Dohlen}, Kjetil and {dos Santos}, Jesulino B. and {dos Santos}, Leandro H. and {Ellis}, Richard and {Fabricius}, Maximilian and {Ferrand}, Didier and {Ferreira}, D{\'e}cio and {Golebiowski}, Mirek and {Greene}, Jenny E. and {Gross}, Johannes and {Gunn}, James E. and {Hammond}, Randolph and {Harding}, Albert and {Hart}, Murdock and {Heckman}, Timothy M. and {Hirata}, Christopher M. and {Ho}, Paul and {Hope}, Stephen C. and {Hovland}, Larry and {Hsu}, Shu-Fu and {Hu}, Yen-Shan and {Huang}, Ping-Jie and {Jaquet}, Marc and {Jing}, Yipeng and {Karr}, Jennifer and {Kimura}, Masahiko and {King}, Matthew E. and {Komatsu}, Eiichiro and {Le Brun}, Vincent and {Le F{\`e}vre}, Olivier and {Le Fur}, Arnaud and {Le Mignant}, David and {Ling}, Hung-Hsu and {Loomis}, Craig P. and {Lupton}, Robert H. and {Madec}, Fabrice and {Mao}, Peter and {Marrara}, Lucas S. and {Mendes de Oliveira}, Claudia and {Minowa}, Yosuke and {Morantz}, Chaz and {Murayama}, Hitoshi and {Murray}, Graham J. and {Ohyama}, Youichi and {Orndorff}, Joseph and {Pascal}, Sandrine and {Pereira}, Jefferson M. and {Reiley}, Daniel and {Reinecke}, Martin and {Ritter}, Andreas and {Roberts}, Mitsuko and {Schwochert}, Mark A. and {Seiffert}, Michael D. and {Smee}, Stephen A. and {Sodre}, Laerte and {Spergel}, David N. and {Steinkraus}, Aaron J. and {Strauss}, Michael A. and {Surace}, Christian and {Suto}, Yasushi and {Suzuki}, Nao and {Swinbank}, John and {Tait}, Philip J. and {Takada}, Masahiro and {Tamura}, Tomonori and {Tanaka}, Yoko and {Tresse}, Laurence and {Verducci}, Orlando and {Vibert}, Didier and {Vidal}, Clement and {Wang}, Shiang-Yu and {Wen}, Chih-Yi and {Yan}, Chi-Hung and {Yasuda}, Naoki},
        title = "{Prime Focus Spectrograph (PFS) for the Subaru telescope: overview, recent progress, and future perspectives}",
    booktitle = {Ground-based and Airborne Instrumentation for Astronomy VI},
         year = 2016,
       editor = {{Evans}, Christopher J. and {Simard}, Luc and {Takami}, Hideki},
       series = {Society of Photo-Optical Instrumentation Engineers (SPIE) Conference Series},
       volume = {9908},
        month = aug,
          eid = {99081M},
        pages = {99081M},
          doi = {10.1117/12.2232103},
archivePrefix = {arXiv},
       eprint = {1608.01075},
 primaryClass = {astro-ph.IM},
       adsurl = {https://ui.adsabs.harvard.edu/abs/2016SPIE.9908E..1MT}
}

@ARTICLE{Zamora2023,
       author = {{Zamora}, S. and {D{\'\i}az}, A.~I.},
        title = "{Revising the cross correlation technique at high spectral resolution}",
      journal = {arXiv e-prints},
         year = 2023,
        month = oct,
          eid = {arXiv:2310.04133},
        pages = {arXiv:2310.04133},
          doi = {10.48550/arXiv.2310.04133},
archivePrefix = {arXiv},
       eprint = {2310.04133},
 primaryClass = {astro-ph.GA},
       adsurl = {https://ui.adsabs.harvard.edu/abs/2023arXiv231004133Z}
}

@article{Heavens1993,
    author = {Heavens, A. F.},
    title = {Galaxy redshifts: improved techniques},
    journal = {Monthly Notices of the Royal Astronomical Society},
    volume = {263},
    number = {3},
    pages = {735-741},
    year = {1993},
    month = {08},
    issn = {0035-8711},
    doi = {10.1093/mnras/263.3.735},
    url = {https://doi.org/10.1093/mnras/263.3.735},
    eprint = {https://academic.oup.com/mnras/article-pdf/263/3/735/4039762/mnras263-0735.pdf},
}

@article{Zucker2003,
    author = {Zucker, S.},
    title = {Cross-correlation and maximum-likelihood analysis: a new approach to combining cross-correlation functions},
    journal = {Monthly Notices of the Royal Astronomical Society},
    volume = {342},
    number = {4},
    pages = {1291-1298},
    year = {2003},
    month = {07},
    issn = {0035-8711},
    doi = {10.1046/j.1365-8711.2003.06633.x},
    url = {https://doi.org/10.1046/j.1365-8711.2003.06633.x},
    eprint = {https://academic.oup.com/mnras/article-pdf/342/4/1291/2827082/342-4-1291.pdf},
}

@ARTICLE{Bouchy2001,
       author = {{Bouchy}, F. and {Pepe}, F. and {Queloz}, D.},
        title = "{Fundamental photon noise limit to radial velocity measurements}",
      journal = {\aap},
         year = 2001,
        month = aug,
       volume = {374},
        pages = {733-739},
          doi = {10.1051/0004-6361:20010730},
       adsurl = {https://ui.adsabs.harvard.edu/abs/2001A&A...374..733B}
}

@ARTICLE{Tranin2025,
       author = {{Tranin}, Hugo and {Blagorodnova}, Nadejda and {G{\'o}mez-Mu{\~n}oz}, Marco A. and {Wavasseur}, Maxime and {Groot}, Paul J. and {Landsberg}, Lloyd and {Stoppa}, Fiorenzo and {Bloemen}, Steven and {Vreeswijk}, Paul M. and {Pieterse}, Dani{\"e}lle L.~A. and {van Roestel}, Jan and {Scaringi}, Simone and {Faris}, Sara},
        title = "{A catalog to unite them all: REGALADE, a revised galaxy compilation for the advanced detector era}",
      journal = {arXiv e-prints},
         year = 2025,
        month = aug,
          eid = {arXiv:2508.13267},
        pages = {arXiv:2508.13267},
          doi = {10.48550/arXiv.2508.13267},
archivePrefix = {arXiv},
       eprint = {2508.13267},
 primaryClass = {astro-ph.GA},
       adsurl = {https://ui.adsabs.harvard.edu/abs/2025arXiv250813267T}
}

@article{Blagorodnova2018,
doi = {10.1088/1538-3873/aaa53f},
url = {https://doi.org/10.1088/1538-3873/aaa53f},
year = {2018},
month = {feb},
publisher = {The Astronomical Society of the Pacific},
volume = {130},
number = {985},
pages = {035003},
author = {Blagorodnova, Nadejda and Neill, James D. and Walters, Richard and Kulkarni, Shrinivas R. and Fremling, Christoffer and Ben-Ami, Sagi and Dekany, Richard G. and Fucik, Jason R. and Konidaris, Nick and Nash, Reston and Ngeow, Chow-Choong and Ofek, Eran O. and Sullivan, Donal O’ and Quimby, Robert and Ritter, Andreas and Vyhmeister, Karl E.},
title = {The SED Machine: A Robotic Spectrograph for Fast Transient Classification},
journal = {Publications of the Astronomical Society of the Pacific}
}

@ARTICLE{Vernet2011,
       author = {{Vernet}, J. and {Dekker}, H. and {D'Odorico}, S. and {Kaper}, L. and {Kjaergaard}, P. and {Hammer}, F. and {Randich}, S. and {Zerbi}, F. and {Groot}, P.~J. and {Hjorth}, J. and {Guinouard}, I. and {Navarro}, R. and {Adolfse}, T. and {Albers}, P.~W. and {Amans}, J.-P. and {Andersen}, J.~J. and {Andersen}, M.~I. and {Binetruy}, P. and {Bristow}, P. and {Castillo}, R. and {Chemla}, F. and {Christensen}, L. and {Conconi}, P. and {Conzelmann}, R. and {Dam}, J. and {de Caprio}, V. and {de Ugarte Postigo}, A. and {Delabre}, B. and {di Marcantonio}, P. and {Downing}, M. and {Elswijk}, E. and {Finger}, G. and {Fischer}, G. and {Flores}, H. and {Fran{\c{c}}ois}, P. and {Goldoni}, P. and {Guglielmi}, L. and {Haigron}, R. and {Hanenburg}, H. and {Hendriks}, I. and {Horrobin}, M. and {Horville}, D. and {Jessen}, N.~C. and {Kerber}, F. and {Kern}, L. and {Kiekebusch}, M. and {Kleszcz}, P. and {Klougart}, J. and {Kragt}, J. and {Larsen}, H.~H. and {Lizon}, J.-L. and {Lucuix}, C. and {Mainieri}, V. and {Manuputy}, R. and {Martayan}, C. and {Mason}, E. and {Mazzoleni}, R. and {Michaelsen}, N. and {Modigliani}, A. and {Moehler}, S. and {M{\o}ller}, P. and {Norup S{\o}rensen}, A. and {N{\o}rregaard}, P. and {P{\'e}roux}, C. and {Patat}, F. and {Pena}, E. and {Pragt}, J. and {Reinero}, C. and {Rigal}, F. and {Riva}, M. and {Roelfsema}, R. and {Royer}, F. and {Sacco}, G. and {Santin}, P. and {Schoenmaker}, T. and {Spano}, P. and {Sweers}, E. and {Ter Horst}, R. and {Tintori}, M. and {Tromp}, N. and {van Dael}, P. and {van der Vliet}, H. and {Venema}, L. and {Vidali}, M. and {Vinther}, J. and {Vola}, P. and {Winters}, R. and {Wistisen}, D. and {Wulterkens}, G. and {Zacchei}, A.},
        title = "{X-shooter, the new wide band intermediate resolution spectrograph at the ESO Very Large Telescope}",
      journal = {\aap},
         year = 2011,
        month = dec,
       volume = {536},
          eid = {A105},
        pages = {A105},
          doi = {10.1051/0004-6361/201117752},
archivePrefix = {arXiv},
       eprint = {1110.1944},
 primaryClass = {astro-ph.IM},
       adsurl = {https://ui.adsabs.harvard.edu/abs/2011A&A...536A.105V}
}

@ARTICLE{Hook2004,
       author = {{Hook}, I.~M. and {J{\o}rgensen}, Inger and {Allington-Smith}, J.~R. and {Davies}, R.~L. and {Metcalfe}, N. and {Murowinski}, R.~G. and {Crampton}, D.},
        title = "{The Gemini-North Multi-Object Spectrograph: Performance in Imaging, Long-Slit, and Multi-Object Spectroscopic Modes}",
      journal = {\pasp},
         year = 2004,
        month = may,
       volume = {116},
       number = {819},
        pages = {425-440},
          doi = {10.1086/383624},
       adsurl = {https://ui.adsabs.harvard.edu/abs/2004PASP..116..425H}
}

@INPROCEEDINGS{Djupvik2010,
       author = {{Djupvik}, Anlaug Amanda and {Andersen}, Johannes},
        title = "{The Nordic Optical Telescope}",
    booktitle = {Highlights of Spanish Astrophysics V},
         year = 2010,
       editor = {{Diego}, Jose M. and {Goicoechea}, Luis J. and {Gonz{\'a}lez-Serrano}, J. Ignacio and {Gorgas}, Javier},
       series = {Astrophysics and Space Science Proceedings},
       volume = {14},
        month = jan,
        pages = {211},
          doi = {10.1007/978-3-642-11250-8_21},
archivePrefix = {arXiv},
       eprint = {0901.4015},
 primaryClass = {astro-ph.IM},
       adsurl = {https://ui.adsabs.harvard.edu/abs/2010ASSP...14..211D}
}

@INPROCEEDINGS{Piascik2014,
       author = {{Piascik}, A.~S. and {Steele}, Iain A. and {Bates}, Stuart D. and {Mottram}, Christopher J. and {Smith}, R.~J. and {Barnsley}, R.~M. and {Bolton}, B.},
        title = "{SPRAT: Spectrograph for the Rapid Acquisition of Transients}",
    booktitle = {Ground-based and Airborne Instrumentation for Astronomy V},
         year = 2014,
       editor = {{Ramsay}, Suzanne K. and {McLean}, Ian S. and {Takami}, Hideki},
       series = {Society of Photo-Optical Instrumentation Engineers (SPIE) Conference Series},
       volume = {9147},
        month = jul,
          eid = {91478H},
        pages = {91478H},
          doi = {10.1117/12.2055117},
       adsurl = {https://ui.adsabs.harvard.edu/abs/2014SPIE.9147E..8HP}
}

@ARTICLE{Brown2013,
       author = {{Brown}, T.~M. and {Baliber}, N. and {Bianco}, F.~B. and {Bowman}, M. and {Burleson}, B. and {Conway}, P. and {Crellin}, M. and {Depagne}, {\'E}. and {De Vera}, J. and {Dilday}, B. and {Dragomir}, D. and {Dubberley}, M. and {Eastman}, J.~D. and {Elphick}, M. and {Falarski}, M. and {Foale}, S. and {Ford}, M. and {Fulton}, B.~J. and {Garza}, J. and {Gomez}, E.~L. and {Graham}, M. and {Greene}, R. and {Haldeman}, B. and {Hawkins}, E. and {Haworth}, B. and {Haynes}, R. and {Hidas}, M. and {Hjelstrom}, A.~E. and {Howell}, D.~A. and {Hygelund}, J. and {Lister}, T.~A. and {Lobdill}, R. and {Martinez}, J. and {Mullins}, D.~S. and {Norbury}, M. and {Parrent}, J. and {Paulson}, R. and {Petry}, D.~L. and {Pickles}, A. and {Posner}, V. and {Rosing}, W.~E. and {Ross}, R. and {Sand}, D.~J. and {Saunders}, E.~S. and {Shobbrook}, J. and {Shporer}, A. and {Street}, R.~A. and {Thomas}, D. and {Tsapras}, Y. and {Tufts}, J.~R. and {Valenti}, S. and {Vander Horst}, K. and {Walker}, Z. and {White}, G. and {Willis}, M.},
        title = "{Las Cumbres Observatory Global Telescope Network}",
      journal = {\pasp},
         year = 2013,
        month = sep,
       volume = {125},
       number = {931},
        pages = {1031},
          doi = {10.1086/673168},
archivePrefix = {arXiv},
       eprint = {1305.2437},
 primaryClass = {astro-ph.IM},
       adsurl = {https://ui.adsabs.harvard.edu/abs/2013PASP..125.1031B}
}

@article{Magill2025,
   title={Super-SNID: An Expanded Set of SNID Classes and Templates for the New Era of Wide-field Surveys},
   volume={9},
   ISSN={2515-5172},
   url={http://dx.doi.org/10.3847/2515-5172/adcab6},
   DOI={10.3847/2515-5172/adcab6},
   number={4},
   journal={Research Notes of the AAS},
   publisher={American Astronomical Society},
   author={Magill, Dylan and Fulton, Michael D. and Nicholl, Matt and Smartt, Stephen J. and Angus, Charlotte R. and Srivastav, Shubham and Smith, Ken W.},
   year={2025},
   month=apr, pages={78} }

@ARTICLE{Liu2014,
	author = {{Liu}, Y. and {Modjaz}, M.},
 	title = "{SuperNova IDentification spectral templates of 70 stripped-envelope core-collapse supernovae}",
journal = {ArXiv e-prints},
archivePrefix = "arXiv",
 	eprint = {1405.1437},
primaryClass = "astro-ph.SR",
   	year = 2014,
 	month = may,
	adsurl = {http://adsabs.harvard.edu/abs/2014arXiv1405.1437L}
}

@ARTICLE{Liu2016,
author = {{Liu}, Y.-Q. and {Modjaz}, M. and {Bianco}, F.~B. and {Graur}, O.
},
	title = "{Analyzing the Largest Spectroscopic Data Set of Stripped Supernovae to Improve Their Identifications and Constrain Their Progenitors}",
journal = {\apj},
archivePrefix = "arXiv",
eprint = {1510.08049},
primaryClass = "astro-ph.HE",
 	year = 2016,
	month = aug,
volume = 827,
  	eid = {90},
	pages = {90},
  	doi = {10.3847/0004-637X/827/2/90},
adsurl = {http://adsabs.harvard.edu/abs/2016ApJ...827...90L}
}

@ARTICLE{Modjaz2016,
author = {{Modjaz}, M. and {Liu}, Y.~Q. and {Bianco}, F.~B. and {Graur}, O.
},
	title = "{The Spectral SN-GRB Connection: Systematic Spectral Comparisons between Type Ic Supernovae and Broad-lined Type Ic Supernovae with and without Gamma-Ray Bursts}",
journal = {\apj},
archivePrefix = "arXiv",
eprint = {1509.07124},
primaryClass = "astro-ph.HE",
 	year = 2016,
	month = dec,
volume = 832,
  	eid = {108},
	pages = {108},
  	doi = {10.3847/0004-637X/832/2/108},
adsurl = {http://adsabs.harvard.edu/abs/2016ApJ...832..108M}
}

@ARTICLE{Liu2017,
author = {{Liu}, Y.-Q. and {Modjaz}, M. and {Bianco}, F.~B.},
title = "{Analyzing the Largest Spectroscopic Data Set of Hydrogen-poor Super-luminous Supernovae}",
journal = {\apj},
archivePrefix = "arXiv",
eprint = {1612.07321},
primaryClass = "astro-ph.HE",
year = 2017,
month = aug,
volume = 845,
eid = {85},
pages = {85},
doi = {10.3847/1538-4357/aa7f74},
adsurl = {http://adsabs.harvard.edu/abs/2017ApJ...845...85L}
}

@ARTICLE{Williamson2019,
     author = {{Williamson}, Marc and {Modjaz}, Maryam and {Bianco}, Federica B.},
     title = "{Optimal Classification and Outlier Detection for Stripped-envelope Core-collapse Supernovae}",
     journal = {\apjl},
     year = 2019,
     month = aug,
     volume = {880},
     number = {2},
     eid = {L22},
     pages = {L22},
     doi = {10.3847/2041-8213/ab2edb},
     archivePrefix = {arXiv},
     eprint = {1903.06815},
     primaryClass = {astro-ph.SR},
     adsurl = {https://ui.adsabs.harvard.edu/abs/2019ApJ...880L..22W}
     }

@ARTICLE{Williamson2023,
    author = {{Williamson}, Marc and {Vogl}, Christian and {Modjaz}, Maryam and {Kerzendorf}, Wolfgang and {Singhal}, Jaladh and {Boland}, Teresa and {Burke}, Jamison and {Chen}, Zhihao and {Hiramatsu}, Daichi and {Galbany}, Llu{\'\i}s and {Gonzalez}, Estefania Padilla and {Howell}, D. Andrew and {Jha}, Saurabh W. and {Kwok}, Lindsey A. and {McCully}, Curtis and {Newsome}, Megan and {Pellegrino}, Craig and {Rho}, Jeonghee and {Terreran}, Giacomo and {Wang}, Xiaofeng},
    title = "{SN 2019ewu: A Peculiar Supernova with Early Strong Carbon and Weak Oxygen Features from a New Sample of Young SN Ic Spectra}",
    journal = {\apjl},
    year = 2023,
    month = feb,
    volume = {944},
    number = {2},
    eid = {L49},
    pages = {L49},
    doi = {10.3847/2041-8213/acb549},
    archivePrefix = {arXiv},
    eprint = {2211.04482},
    primaryClass = {astro-ph.HE},
    adsurl = {https://ui.adsabs.harvard.edu/abs/2023ApJ...944L..49W}
}

@ARTICLE{Yesmin2024,
       author = {{Yesmin}, N. and {Pellegrino}, C. and {Modjaz}, M. and {Baer-Way}, R. and {Howell}, D.~A. and {Arcavi}, I. and {Farah}, J. and {Hiramatsu}, D. and {Hosseinzadeh}, G. and {McCully}, C. and {Newsome}, M. and {Padilla Gonzalez}, E. and {Terreran}, G. and {Jha}, S.},
        title = "{Spectral Dataset of Young Type Ib Supernovae and their Time-evolution}",
      journal = {arXiv e-prints},
         year = 2024,
        month = sep,
          eid = {arXiv:2409.04522},
        pages = {arXiv:2409.04522},
          doi = {10.48550/arXiv.2409.04522},
archivePrefix = {arXiv},
       eprint = {2409.04522},
 primaryClass = {astro-ph.HE},
       adsurl = {https://ui.adsabs.harvard.edu/abs/2024arXiv240904522Y}
}

@article{Gupta2016,
   title={HOST GALAXY IDENTIFICATION FOR SUPERNOVA SURVEYS},
   volume={152},
   ISSN={1538-3881},
   url={http://dx.doi.org/10.3847/0004-6256/152/6/154},
   DOI={10.3847/0004-6256/152/6/154},
   number={6},
   journal={The Astronomical Journal},
   publisher={American Astronomical Society},
   author={Gupta, Ravi R. and Kuhlmann, Steve and Kovacs, Eve and Spinka, Harold and Kessler, Richard and Goldstein, Daniel A. and Liotine, Camille and Pomian, Katarzyna and D’Andrea, Chris B. and Sullivan, Mark and Carretero, Jorge and Castander, Francisco J. and Nichol, Robert C. and Finley, David A. and Fischer, John A. and Foley, Ryan J. and Kim, Alex G. and Papadopoulos, Andreas and Sako, Masao and Scolnic, Daniel M. and Smith, Mathew and Tucker, Brad E. and Uddin, Syed and Wolf, Rachel C. and Yuan, Fang and Abbott, Tim M. C. and Abdalla, Filipe B. and Benoit-Lévy, Aurélien and Bertin, Emmanuel and Brooks, David and Rosell, Aurelio Carnero and Kind, Matias Carrasco and Cunha, Carlos E. and Costa, Luiz N. da and Desai, Shantanu and Doel, Peter and Eifler, Tim F. and Evrard, August E. and Flaugher, Brenna and Fosalba, Pablo and Gaztañaga, Enrique and Gruen, Daniel and Gruendl, Robert and James, David J. and Kuehn, Kyler and Kuropatkin, Nikolay and Maia, Marcio A. G. and Marshall, Jennifer L. and Miquel, Ramon and Plazas, Andrés A. and Romer, A. Kathy and Sánchez, Eusebio and Schubnell, Michael and Sevilla-Noarbe, Ignacio and Sobreira, Flávia and Suchyta, Eric and Swanson, Molly E. C. and Tarle, Gregory and Walker, Alistair R. and Wester, William},
   year={2016},
   month=nov, pages={154} }

@inproceedings{McLachlan2000,
  title={Finite Mixture Models},
  author={Geoffrey J. McLachlan and David Peel},
  booktitle={Wiley Series in Probability and Statistics},
  year={2000},
  url={https://api.semanticscholar.org/CorpusID:124985575}
}



\appendix

\section{Template Manager}
\label{appendix:template-manager}

\textsc{SNID--SAGE} includes a dedicated Template Manager, accessible via the \texttt{snid-sage-templates} command, for inspecting, curating, and extending the spectral template library. The tool serves both as a quality-control environment for the distributed templates and as a workflow for incorporating user-contributed spectra in a controlled and reproducible manner.

\begin{figure}
\centering
\includegraphics[width=\linewidth]{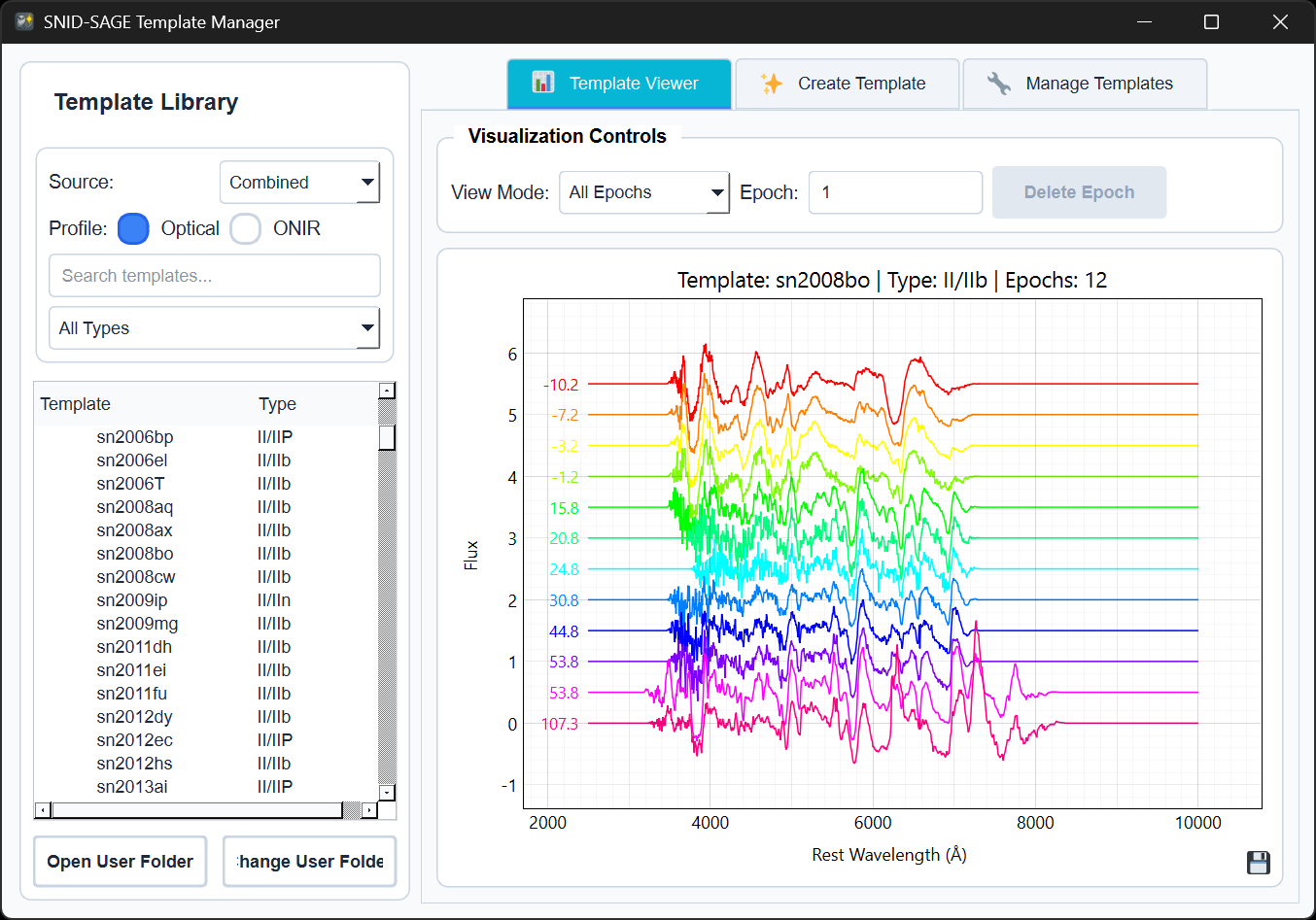}
\caption{\textbf{Template Manager interface.} Example view of a multi-epoch template (SN\,2008bo, Type~IIb) with phase-coded spectra and interactive inspection tools.}
\label{fig:template_manager}
\end{figure}

The Template Manager allows interactive inspection of template spectra across multiple epochs, including visual comparison, metadata editing, and verification of subtype, phase, and redshift assignments (Figure~\ref{fig:template_manager}). Users can create new templates from input spectra using the same preprocessing procedures employed in the main classification pipeline, ensuring full consistency between template construction and analysis.

User-created templates are maintained separately from the distributed base library and are merged at runtime, allowing corrections or extensions without modifying the original dataset.

\section{Example of AI-assisted summary}
\label{appendix:ai-summary}

As described in Section~\ref{sec:methods:gui-ai}, the \textsc{SNID--SAGE} graphical interface includes an optional feature that generates a structured natural-language summary of the classification results using an externally configured large language model (LLM). This functionality is intended purely for reporting and human-readable interpretation; it does not influence any part of the numerical analysis.

The LLM is provided with a formatted summary of the \textsc{SNID--SAGE} output, including the best-fit type and subtype, redshift and phase estimates with uncertainties, match-quality and confidence metrics, and a subset of the highest-ranked template matches. An example output, representative of that produced using a publicly available model via OpenRouter, is shown below.

\begin{mdframed}[backgroundcolor=gray!10, linewidth=0.4pt]
\small
\textbf{SNID AI ANALYSIS SUMMARY}

\noindent
The spectrum is classified as a Type II supernova of subtype IIP, best matched by the template \texttt{sn1999em} (II/IIP), with High match quality and High confidence relative to the next-best alternative. The cluster-weighted redshift is $z = 0.0180 \pm 0.0021$, and the inferred phase is approximately 2--3 days relative to maximum light, with an uncertainty of about 11 days. The consistency of the top-ranked template matches supports the robustness of the classification, although the phase estimate remains moderately uncertain.
\end{mdframed}

The exact wording of the generated summary will vary depending on the chosen model and prompt configuration. The summaries are intended to facilitate rapid reporting and interpretation, particularly in time-domain follow-up settings, and should be regarded as descriptive outputs derived from the deterministic \textsc{SNID--SAGE} analysis rather than as independent classifications.


\bsp	
\label{lastpage}
\end{document}